\documentclass[10pt, twocolumn, final, journal,  letterpaper]{IEEEtran}
\usepackage{graphicx}
\usepackage{epstopdf}
\usepackage{amsmath}
\usepackage{amssymb}
\usepackage{cite}
\usepackage{color}
\usepackage{amsthm}
\theoremstyle{plain}	 
\newtheorem{thm}{Theorem}
\newtheorem{lem}{Lemma}
\newtheorem{cor}{Corollary}[lem]
\usepackage{mathtools}
\usepackage[normalem]{ulem}
\usepackage{cancel}

\def\th{{\rm th}}
\def\by{\mathbf{y}}
\def\bx{\mathbf{x}}

\def\bZ{\mathbf{Z}}

\def\bX{\mathbf{X}}
\def\bof{\mathbf{f}}

\def\bg{\mathbf{g}}
\def\FA{{\rm FA}}
\def\MD{{\rm MD}}
\def\cX{{\cal X}}

\newcommand{\R}{\mathbb{R}}

\newcommand{\Prob}{\mathbb{P}}
\newcommand{\done}{\hspace*{\fill} \rule{1.8mm}{2.5mm}}

\begin{document}
\title{Covert Communication in the Presence of an Uninformed Jammer}
\author{Tamara~V.~Sobers,~\IEEEmembership{Member,~IEEE,
} Boulat~A.~Bash,~\IEEEmembership{Member,~IEEE,}
Saikat~Guha,~\IEEEmembership{Senior~Member,~IEEE,}
Don~Towsley,~\IEEEmembership{Fellow, IEEE,} and
Dennis~Goeckel,~\IEEEmembership{Fellow,~IEEE}
\thanks{This work was sponsored by the National Science Foundation under grants ECCS-1309573 and CNS-1564067, and DARPA under contract number HR0011-16-C-0111.}
\thanks{T. V. Sobers was with the Electrical and Computer Engineering Department, University of Massachusetts, Amherst, Massachusetts. She is now with The MITRE Corporation, Bedford, Massachusetts (e-mail: tsobers@mitre.org).}
\thanks{B. A. Bash is with Raytheon BBN, Cambridge, Massachusetts (e-mail: boulat.bash@raytheon.com).}
\thanks{ S. Guha was with Raytheon BBN,  Cambridge, Massachusetts.  He is now with the College of Optical Sciences, University of Arizona, Tuscon, Arizona (e-mail: saikat@email.arizona.edu).}
\thanks{D. Towsley is with the College of Information and Computer Sciences, University of Massachusetts, Amherst, Massachusetts (e-mail: towsley@cs.umass.edu).}
\thanks{D. Goeckel is with the Electrical and Computer Engineering Department, University of Massachusetts, Amherst, Massachusetts (e-mail: goeckel@ecs.umass.edu). }
}
 


\IEEEoverridecommandlockouts
\IEEEpubid{\makebox[\columnwidth]{
\copyright2017
IEEE \hfill} \hspace{\columnsep}\makebox[\columnwidth]{ }} 

\maketitle
\begin{abstract}
Recent work has established that when transmitter Alice wishes to communicate reliably to recipient Bob without detection by warden Willie, with additive white Gaussian noise (AWGN) channels between all parties, communication is limited to $\mathcal{O}(\sqrt{n})$ bits in $n$ channel uses.  However, this assumes Willie has an accurate statistical characterization of the channel.  When Willie has uncertainty about such and his receiver is limited to a threshold test on the received power, Alice can transmit covertly with a power that does not decrease with $n$, thus conveying $\mathcal{O}(n)$  bits covertly and reliably in $n$ uses of an AWGN channel.  Here, we consider covert communication of $\mathcal{O}(n)$ bits in $n$ channel uses while generalizing the environment and removing any restrictions on Willie's receiver.   We assume an uninformed ``jammer'' is present to help Alice, and we consider AWGN and block fading channels.  In some scenarios, Willie's optimal detector is a threshold test on the received power. When the channel between the jammer and Willie has multiple fading blocks per codeword, a threshold test on the received power is not optimal.  However, we establish that Alice can remain covert with a transmit power that does not decrease with $n$ even when Willie employs an optimal detector.
\end{abstract}


\section{Introduction}
Much of secure communications centers on preventing an adversary from determining the content of the message. However, there are circumstances where communicating parties Alice and Bob may want \textit {covert} communication: hiding the very existence of their communication from a watchful adversary Willie.   Examples include communicating in the presence of an authoritarian government who may want to curtail any organization by certain entities, or military communications where detection might inform an adversary that there is activity in a given geographical area.  

As defined precisely below, recent work has studied reliable covert communication, which requires: (i) Willie's error in detecting that Alice transmitted a message to Bob be arbitrarily close to random guessing; and (ii) Bob's error of recovering Alice's message be arbitrarily small.  When the Alice-to-Bob and Alice-to-Willie channels are additive white Gaussian noise (AWGN) channels, \cite{bash_isit2012} and \cite{bash_jsac} showed a \emph{square root law} (SRL): provided Alice and Bob share a secret of sufficient length prior to transmission, Alice can communicate covertly to Bob if and only if she employs a per-symbol power of no more than
 $\mathcal{O}(1/\sqrt{n})$, which decreases to 0 in the limit of large $n$.   Thus, $\mathcal{O}(\sqrt{n})$ bits (and no more) can be transmitted in $n$ channel uses \cite{bash_jsac}.   Follow-on work has considered the length of the pre-shared secret in \cite{jaggi_isit} and \cite{bloch15covert}, characterization of the constant hidden by Big-$\mathcal{O}$ notation in \cite{bloch15covert} and \cite{wang_2016}, and both the theory and experimental verification of covert communication over quantum channels in \cite{bash_nature} and \cite{azadeh_isit_2016}. 

Subsequent work considered whether positive rate covert communications, which requires the transmission of $O(n)$ bits in $n$ channel uses, is possible.  Lee \textit{et al.}~in \cite{lee15posratecovertjstsp} demonstrated that positive rate is indeed achievable over AWGN channels if Willie has uncertainty about the statistics of the background noise and is restricted to a receiver that employs a threshold on the received power when attempting to detect Alice.   Che \textit{et al.}~in \cite{jaggi_itw} proved that positive rate is achievable if Willie has uncertainty in the parameters of the binary symmetric channel between Alice and himself.    In \cite{goeckel_commL}, the authors re-visit the results of \cite{lee15posratecovertjstsp} and \cite{jaggi_itw}. Rather than starting with parametric uncertainty in Willie's knowledge of the noise statistics, \cite{goeckel_commL} allows Willie to have access to a large collection of inputs spanning many possible codeword slots and to employ them in any way that he deems suitable.  Then, the lack of knowledge of channel statistics at Willie does not increase the order of the covert throughput from Alice to Bob \cite{goeckel_commL}.  This is because Willie is able to use any ``quiet'' periods to estimate the noise statistics of his receiver accurately and then detect if Alice is transmitting, even if he does not know a priori the time at which Alice might transmit.

In this work, we allow Willie to have a general receiver, as in \cite{goeckel_commL}, but we seek conditions under which Alice can transmit with power not decreasing in the blocklength $n$; in the case of an AWGN channel between Alice and Bob, this then achieves the transmission of $O(n)$ bits covertly in $n$ channel uses.  To do such, we add another node to the environment, the ``jammer'', who Willie knows is transmitting.   For example, this might be a jammer in an electronic warfare (EW) environment placed by Alice and Bob, or, as discussed in Section VI, a jammer placed in the environment by Willie for other security objectives.  If this jammer randomly varies his/her transmit power appropriately or if time-varying multipath fading causes sufficient variation, channel estimation during periods outside the time period when Willie is attempting to detect Alice's transmission cannot be used to estimate the statistics of the noise impacting Willie's receiver during the period of interest.  Hence, the results of \cite{goeckel_commL} do not apply; rather, we arrive at a similar mathematical problem to that considered in \cite{lee15posratecovertjstsp}.   A limitation of the achievability results of \cite{lee15posratecovertjstsp} is that the power detector is not established to be the optimal receiver for Willie; in fact, in the case of block fading channels with multiple fading blocks per codeword, it is known to be sub-optimal.  Here, in contrast to \cite{lee15posratecovertjstsp}, we establish covert communication against any detector that Willie might employ.  

We consider both additive white Gaussian noise (AWGN) and standard block fading channels.  Note that the problem is readily solved if the jammer and Alice are closely coordinated (i.e.~, an ``informed'' jammer) by the following construction.  Alice generates a codebook by drawing codeword symbols independently from a Gaussian distribution, and provides this codebook only to Bob as the shared secret.   At the time Alice starts to transmit a codeword, the jammer turns down the power of his transmission of Gaussian noise, and then he turns it back up at the moment Alice finishes transmitting.  Willie is then unable to determine that any change has taken place when Alice is transmitting.   We are interested in the case where the jammer and Alice do not coordinate.   In the AWGN case, our construction has the jammer randomly change his/her power of the Gaussian noise in each ``slot'' of $n$ symbols, where $n$ is the codeword length used by Alice.  By doing such, Willie is unaware of the background noise to expect and it is plausible, particularly based on the work of \cite{lee15posratecovertjstsp}, that Alice should be able to achieve positive rate covert communication to Bob.  To establish this result rigorously against an arbitrary receiver at Willie, we first establish that Willie's optimal receiver is indeed a comparison of the received power to a threshold, from which the achievability of positive rate covert communication follows.  

We then consider a block fading channel with $M$ fading blocks per codeword of length $n$.   If $M=1$, we demonstrate that a threshold test on the total received power in the codeword slot is the optimal detector at Willie, from which covert transmission by Alice with power not decreasing in the blocklength $n$ follows.  When $M > 1$, a threshold test on the total received power at Willie is sub-optimal.  Thus, we first establish a technical property on the structure of Willie's optimal detector and then show that this property suffices to establish the ultimate goal when the jammer-to-Willie channel is an $M > 1$ block fading channel:  Alice can covertly transmit with a power that does not decrease with her blocklength $n$.

\noindent Our \textit{main contributions} are:
\begin{enumerate}
\item The consideration of covert communication in the presence of an uninformed jammer.
\item The demonstration of the optimality of a power detector at Willie for the AWGN and $M=1$ block fading cases, from which the ability of Alice to transmit covertly with a power that does not decrease with her blocklength follows.
\item The demonstration of the ability for Alice to transmit covertly with a power that does not decrease with her blocklength in the $M>1$ block fading scenario, even when Willie uses an optimal detector (which is not a power detector in this case).
\end{enumerate}

Section II presents the system model and performance metrics considered in this work.  Section III considers the AWGN case, and Section IV extends these results to the mathematically similar $M=1$ block fading case.   The $M>1$ block fading case requires a significantly different approach, which is described in Section V.  
Section VI summarizes two potential points of discussion based on the results presented: 1) in the electronic warfare model, active jamming by adversaries may help facilitate covert communication; and 2) the difference between positive rate communication in the wireless scenarios presented in this work and typical steganography systems.  Finally, Section VII presents conclusions and ideas for continuing work.


\section{System Model and Metrics}
\subsection{System Model}
\label{sec:model}

\begin{figure}[ht]
  \begin{center}
	\includegraphics[scale=.5]{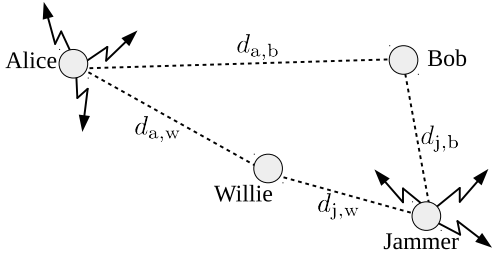}
  \end{center}
  \caption{Wireless communication scenario.  With the help of a jammer, Alice attempts to transmit covertly to Bob in the presence of a watchful adversary Willie.}
  \label{fig:scenario}
\end{figure}

Consider a scenario where Alice (``a") would like to communicate covertly to Bob (``b") without detection by a warden Willie (``w"), and suppose a jammer (``j") is active in the environment who is willing to assist with this communication. 
The geographic model is shown in Figure \ref{fig:scenario}.  The distances from Alice to Willie and Alice to Bob are denoted by $d_{\rm a,w}$ and $d_{\rm a,b}$ respectively.   The distances from the jammer to Willie and the jammer to Bob are $d_{\rm j,w}$ and $d_{\rm j,b}$ respectively. 

\begin{figure*}[ht]
  \begin{center}
	\includegraphics[scale=.4]{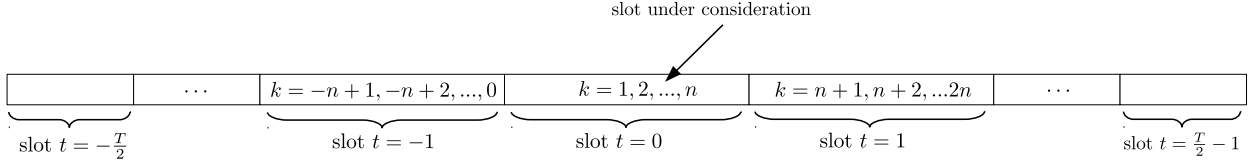}
  \end{center}
  \caption{Representation of the indexing of the $nT$ symbol periods in $T$ slots, each of length $n$.  Alice decides to transmit in slot $t=0$ with probability $p$, and Willie attempts to detect a transmission in that slot.}
  \label{fig:slot_model}
\end{figure*}

We are interested in Alice's ability to transmit covertly in a slot equal to the codeword length $n$ and Willie's ability to detect such a transmission in that slot.   For integer constant $T>0$, we consider a discrete-time channel with $T$ slots, each of length $n$ symbols, as shown in Figure \ref{fig:slot_model}, with the $nT$ symbols indexed by $k=-\frac{T}{2} n+1, -\frac{T}{2}n+2\ldots,-2,-1,0,1,2,\ldots, \frac{T}{2} n -1, \frac{T}{2}n$.  We assume that the slot of interest is slot $t=0$; hence, Alice may (or may not) transmit for a duration of $n$ symbols starting at time $k=1$, and Willie's goal is to detect whether or not such a transmission took place using observations for all $k=-\frac{T}{2} n+1, -\frac{T}{2}n+2\ldots,-2,-1,0,1,2,\ldots, \frac{T}{2} n -1, \frac{T}{2}n$, since observations outside of $k=1,2,\ldots,n$ might be useful to Willie in estimating aspects of the environment \cite{goeckel_commL}.   The jammer is ``uninformed" in the sense that it does not know if Alice transmits, and if Alice transmits, the jammer does not know that Alice is going to use a slot starting at time $k=1$.


Alice transmits a message with probability $p$ and if she decides to transmit, she maps her message to the complex symbol sequence $\bof=[f_1,~f_2,~\ldots~,f_n]$ and sends it in the $t=0$ slot corresponding to symbols $k=1,2,\ldots,n$.  The jammer is allowed to transmit continuously (in all symbols of all slots) subject only to an average power limitation of $P_{\rm{max}}$ per symbol.  
Let the (complex) signal transmitted by the jammer for all time slots be given by $\{\bg_t\}_{t=-\frac{T}{2}}^{\frac{T}{2} -1}$, where $\bg_t = [g_{t n+1}, g_{t n + 2}, \ldots, g_{t n + n}]$ is the vector of transmitted jamming signals sent during the $t^{\text{th}}$ slot, with the per symbol power constraint $E[|g_k|^2] \leq P_{\rm{max}}$.  

\subsubsection{AWGN channel model}

\noindent Consider first the AWGN channel.  Denote the collection of channel outputs at Willie over all time slots as:  $\{{\mathbf{Z}}_{t}\}_{t=-\frac{T}{2}}^{\frac{T}{2} -1}$, where ${ \mathbf{Z}}_{t} = [{Z}_{t n+1}, {Z}_{t n+2}, \ldots, { Z}_{t n + n}]$ is the vector of observations collected during the $t^{\text{th}}$ slot.  Hence, for slot $t$, $i=1,2,\ldots, n$:
\begin{align}
{Z}_{t n +i} = \begin{cases}
 \frac{f_i}{d_{\rm a,w}^{\alpha/2}} + \frac{g_{t n + i}}{d_{\rm j,w}^{\alpha/2}} + {N}_{t n + i}^{\rm (w)},&\!\!\mbox{Alice transmits and}~t=0\\
\frac{g_{t  n + i}}{d_{\rm j,w}^{\alpha/2}} + {N}_{t  n + i}^{\rm (w)}, &\!\!\mbox{else,}
\end{cases}\label{eq:w_rx_basic1}
  \end{align}
where $\alpha$ is the path-loss exponent, and $\{N^{\rm (w)}_k, k=-\frac{T}{2} n+1, -\frac{T}{2}n+2\ldots,-2,-1,0,1,2,\ldots, \frac{T}{2} n -1, \frac{T}{2}n\}$ is a set of independent and identically distributed (i.i.d.) zero-mean complex Gaussian random variables, each with variance {$E[|N^{\rm (w)}_k|^2]=\sigma_{\rm w}^2$}.  

Similarly, denote the collection of channel outputs at Bob over all time slots as:
$\{{ \mathbf{Y}}_t\}_{t=-\frac{T}{2}}^{\frac{T}{2} -1}$, where ${ \mathbf{Y}}_t = [{Y}_{t n+1}, {Y}_{t n+2}, \ldots, {Y}_{t n + n}]$ is the vector of observations collected during the $t^{\text{th}}$ slot.  Hence, for slot $t$, $i=1,2,\ldots, n$:
\begin{align}
{Y}_{t n +i} = \begin{cases}
 \frac{f_i}{d_{\rm a,b}^{\alpha/2}} + \frac{g_{t n + i}}{d_{\rm j,b}^{\alpha/2}} + {N_{t n + i}^{\rm (b)}},&\!\!\mbox{Alice transmits and}~t=0\\
\frac{g_{t n + i}}{d_{\rm j,b}^{\alpha/2}} + {N_{t n + i}^{\rm (b)}},&\!\!\mbox{else,}
\end{cases}\label{eq:w_rx_basic2}
  \end{align}
where $\{N^{\rm (b)}_k, k=-\frac{T}{2} n+1, -\frac{T}{2}n+2\ldots,-2,-1,0,1,2,\ldots, \frac{T}{2} n -1, \frac{T}{2}n\}$ is a set of i.i.d.~zero-mean complex Gaussian random variables, each with variance $E[|N^{\rm (b)}_k|^2]=\sigma_{\rm b}^2$.

\subsubsection{Block fading channels}

\begin{figure}[ht]
  \begin{center}
	\includegraphics[scale=.36]{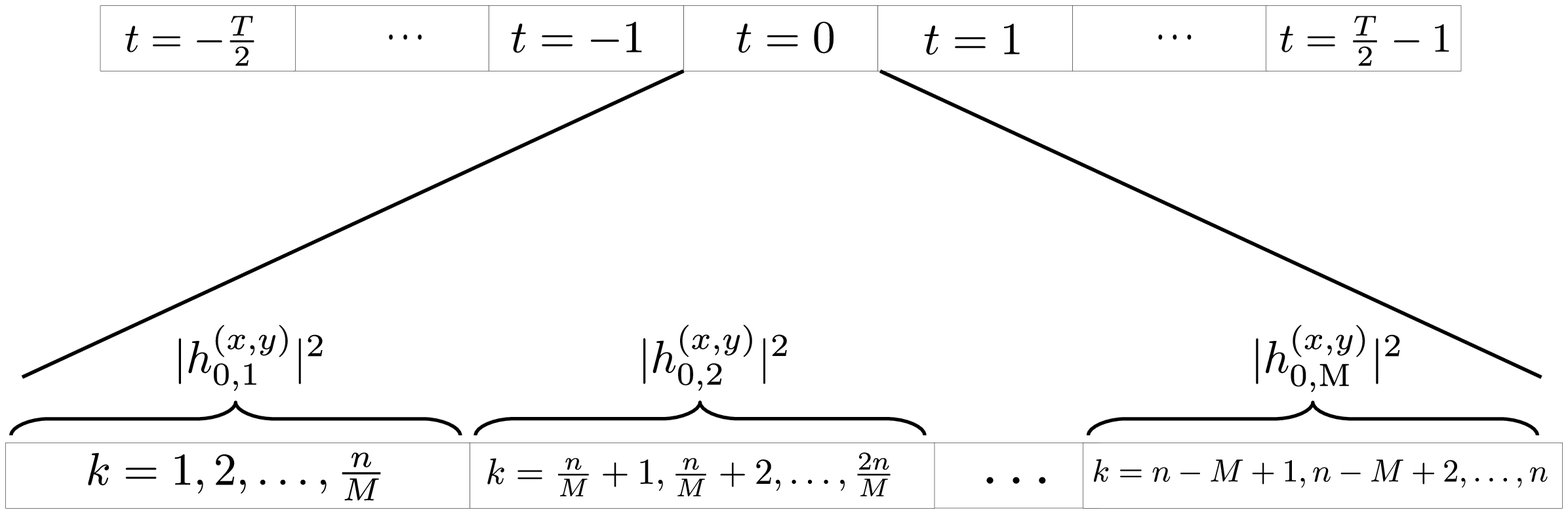}
  \end{center}
  \caption{Model for multiple block fading conditions where $x$ is either Alice or the jammer and $y$ is either Willie or Bob.}
  \label{fig:slot_model_fading}
\end{figure}

Consider next the standard {Rayleigh} block fading channel, as shown in Figure \ref{fig:slot_model_fading}.  The fading is constant for a block of $n/M$ symbols but changes independently to a different value for the next block, where $M$ is the number of fading blocks per codeword slot \cite{tse_vishwanath}.  Denote $h_{t,m}^{(x,y)}$, $m=1,\ldots, M$ as the (complex) fading coefficient for the $m^{\rm th}$ block during slot $t$ between transmitter $x$ and receiver $y$, where $x$ is either ``a" (Alice) or ``j" (jammer), and $y$ is either ``w'' (Willie) or ``b'' (Bob).  By the Rayleigh fading assumption, $h_{t,m}^{(x,y)}$, $m=1,\ldots,M$ is assumed to be a zero mean complex Gaussian random variable with $E[|h_{t,m}^{(x,y)}|^2]=1$ for all channels.  The fading processes affecting different transmitter-receiver pairs are assumed to be independent of each other.
For slot $t$, $i=1,2,\ldots, n$, Willie observes:
\begin{equation}
{Z}_{t n +i}=\begin{cases}
               \frac{h_{t, \lfloor (i-1) \frac{M}{n} \rfloor + 1}^{\rm (a,w)} f_i}{d_{\rm a,w}^{\alpha/2}} &\\
               \quad + \frac{h_{t,\lfloor (i-1) \frac{M}{n} \rfloor + 1}^{\rm (j,w)} g_{t n + i}}{d_{\rm j,w}^{\alpha/2}} +{N}_{t n + i}^{\rm (w)},\!\!\!&\mbox{Alice tx, }t\!=\!0 \\
               \frac{h_{t, \lfloor (i-1) \frac{M}{n} \rfloor + 1}^{\rm (j,w)} g_{t n + i}}{d_{\rm j,w}^{\alpha/2}} + {N}_{t n + i}^{\rm (w)}, &\mbox{else.}
            \end{cases}
\end{equation}
For slot $t$, $i=1,2,\ldots,n$, Bob observes:
\begin{equation}
{Y}_{t n +i} = 
\begin{cases}
 \frac{h_{t, \lfloor (i-1) \frac{M}{n} \rfloor + 1}^{\rm (a,b)} f_i}{d_{\rm a,b}^{\alpha/2}} & \\ \quad + \frac{h_{t,\lfloor (i-1) \frac{M}{n} \rfloor + 1}^{\rm (j,b)} g_{t  n + i}}{d_{\rm j,b}^{\alpha/2}} + {N}_{t  n + i}^{\rm (b)},\!\!\!&\mbox{Alice tx, }t\!=\!0 \\
\frac{h_{t, \lfloor (i-1) \frac{M}{n} \rfloor + 1}^{\rm (j,b)} g_{t n + i}}{d_{\rm j,b}^{\alpha/2}} + {N}_{t n + i}^{\rm (b)}, &\mbox{else.}
\end{cases}\label{eq:w_rx_basic4}
\end{equation}

\subsection{Metrics, hypothesis testing, and likelihood ratio ordering}\label{sec:metrics}

Based on his observations over all time slots, Willie must determine whether Alice transmitted in time slot $t=0$.  The null hypothesis $(H_0)$ is that Alice did not transmit and the alternative hypothesis $(H_1)$ is that that Alice transmitted a message.  Define $P(H_0)=1-p$ as the probability that Alice does not transmit and $P(H_1)=p$ as the probability that Alice transmits in time slot $t=0$, where we assume (pessimistically) that $p$ is known to Willie.  Willie seeks to minimize his probability of error $\mathbb{P}_e= (1-p) \cdot \mathbb{P}_\FA + p \cdot \mathbb{P}_\MD$, where $\mathbb{P}_{\rm MD}$ and $\mathbb{P}_{\rm FA}$ are the probabilities of missed detection and false alarm at Willie, respectively.  Per \cite{bash_ignorance}, $\mathbb{P}_e\geq \min(p,1-p) \cdot (\mathbb{P}_\FA + \mathbb{P}_\MD)$.  Hence, we will say that
Alice achieves \textit{covert} communication if, for any $\epsilon > 0$, $\mathbb{P}_{\rm MD} + \mathbb{P}_{\rm FA} > 1 - \epsilon$ for $n$ sufficiently large.\footnote{This guarantees that Willie's probability of error is within $\epsilon$ of the probability of error  $\min(p,1-p)$ obtained if he ignores his observations and chooses the hypothesis $H_0$ and $H_1$ that was most likely a priori.}   We will say that Alice can transmit covertly with power not decreasing in $n$ if, for any $\epsilon > 0$, there exists $P_{\rm f} > 0$ not dependent on $n$ (but possibly dependent on $\epsilon$) such that, as $n \rightarrow \infty$, a system employing power $P_{\rm f}$ is covert.  Bob should also be capable of  \textit{reliably} decoding Alice's message \cite{bash_jsac}.  Bob can reliably decode messages from Alice if, for any $\delta > 0$, his probability of error is less than $\delta$ for $n$ sufficiently large.   

We assume that Willie has full knowledge of the statistical model: the parameters for Alice's random codebook generation and the jammer's random interference generation, the noise variance $\sigma_{\rm w}^2$, and in the case of fading on the Alice-to-Willie channel or jammer-to-Willie link, the statistics of that fading.  Thus, Willie's test is between two simple hypotheses for Alice's transmission state, and he has complete statistical knowledge of his observations when either hypothesis is true.  Therefore, by applying the Neyman-Pearson (NP) criterion, the optimal test for Willie to minimize his probability of error is the {\em likelihood ratio test (LRT)} \cite[Chapter~3.3]{kay1998_detection},
\begin{equation}
\Lambda(\mathbf{\tilde Z}) = \frac{f_{\mathbf{\tilde Z}|H_1}(\mathbf{\tilde Z}|H_1)}{f_{\mathbf{\tilde Z}|H_0}(\mathbf {\tilde Z}|H_0)}  
\mathop{\gtrless}_{H_0}^{H_1} \gamma,  \label{firsttest}
\end{equation}
where $\gamma =P(H_0)/P(H_1)$, and
$f_{\mathbf{\tilde Z}|H_1}(\cdot|H_1)$ and  $f_{\mathbf{\tilde Z}|H_0}(\cdot|H_0)$ are the probability density functions (pdfs) for Willie's observations over all slots given Alice transmitted in the $t=0$ slot or given Alice did not transmit in the $t=0$ slot, respectively.  As can be inferred by the assumption of a power detector for Willie's receiver in \cite{lee15posratecovertjstsp} and made precise in the proof of Theorem 1 below, a desirable property for the likelihood ratio $\Lambda(\cdot)$ to exhibit is {\em monotonicity}.  In the remainder of this section, we describe an approach for establishing such a property that applies in our context. 

We employ the concept of {\em stochastic ordering} \cite{SS94} to derive the desired monotonicity results in a more streamlined fashion relative to our preliminary work in \cite{sobers_asilomar2015}.  We say that random variable $X$ is smaller than $W$ in the likelihood ratio order (written as $X\leq_{\rm lr} W$) when $f_W(x)/f_X(x)$ is non-decreasing over the union of their supports, where $f_W(x)$ and $f_X(x)$ are their respective probability density functions.  Consider a family of pdfs $\{g_\theta(\cdot), \theta \in \cX\}$ where $\cX$ is a subset of the real line.  Let $X (\theta )$ denote a random variable with density $g_\theta(\cdot)$ for fixed parameter $\theta$.  Let $\Theta$ denote a random variable with support $\cX$ and probability distribution function $F_{\Theta}(\cdot)$; we denote $X(\Theta )$ as the random variable that is the mixture of the random variables $X(\theta)$ under distribution $F_{\Theta}(\theta)$; that is, the probability density function of $X(\Theta)$ is given by:
\begin{align}
f_{X(\Theta )}(x) = \int_{\theta \in \cX}g_{\theta}(x) dF(\theta ) , \quad x\in \R.
\end{align}
We will rely on the following result regarding mixtures of random variables. 

\begin{lem} {[Theorem 1.C.11 in \cite{SS94}]}  \label{lem:lrmixtures}
Consider a family of probability density functions $\{g_\theta(\cdot), \theta \in \cX\}$ with $\cX$ a subset of the real line. Let $\Theta_0$ and $\Theta_1$ denote random variables with support in $\cX$ and probability distribution functions $F_0(\theta)$ and $F_1(\theta)$, respectively.  Let $W_0$ and $W_1$ be random variables such that $W_i =_{\rm d} X(\Theta_i)$, $i=0,1$, (where $=_{\rm d}$ is defined as equality in distribution or law):
\begin{align}
f_{W_i} (x) = \int_{\theta \in \cX} g_\theta (x)  dF_i(\theta ) , \quad i=0,1; x\in \R.
\end{align}

\noindent {If }
\begin{align}
X(\theta ) \le_{\rm lr} X(\theta '),  \quad \theta \le \theta ' \label{eq:lem1_X_lr_X}
\end{align}
{and}
\begin{align}
\Theta_0 \le_{\rm lr} \Theta_1,\label{eq:lem1_Th_lr_Th}
\end{align}

\noindent then 
\begin{align}
W_0 \le_{\rm lr} W_1.\label{eq:lem1_mono_holds}
\end{align}
 
\end{lem}

\section{AWGN Channels}\label{sec:awgn}
We first consider the case of additive white Gaussian noise (AWGN) channels between all nodes, with the slot boundaries between Alice, Willie, and the jammer synchronized, and, as in \cite{bash_jsac}, assume that Alice and Bob share a secret of unlimited length.  We provide a construction for Alice and the jammer, and then demonstrate that this construction makes Willie's optimal detector a power detector.  The transmission of $\mathcal{O}(n)$ bits in $n$ channel uses is then demonstrated.  It is assumed that $d_{\rm a,w}$ and $d_{\rm j,w}$ are known to Alice, although it will be readily apparent that a lower-bound to $d_{\rm a,w}$ and an upper-bound to $d_{\rm j,w}$ are sufficient to establish the results.  

{\em Construction:} We employ random coding arguments and generate $K$ codewords, each of length $n$, by independently drawing symbols from a zero-mean complex Gaussian distribution with variance $P_{\rm f}$, where $P_{\rm f}$ is determined later.  This codebook is revealed to Alice and Bob, is used only once, and comprises the shared secret unknown to Willie (and the jammer).  If Alice decides to transmit in slot $t=0$, she selects the codeword corresponding to her message, sets $f_i$ to the $i^{\rm th}$ symbol of that codeword, and transmits the sequence $f_1, f_2,\ldots, f_n$.   The jammer, with knowledge of the slot boundaries but without knowledge of whether Alice transmits in a given slot (or at all), transmits a symbol drawn independently from a zero-mean complex Gaussian distribution during each symbol period.  However, the variance of this Gaussian distribution is not constant; in particular, during the $t^{\rm th}$ slot, the jammer draws each of its symbols independently from a zero-mean Gaussian distribution with variance $E[|g_{t n +i}|^2]=P^{\rm (j)}_t$, $i=1,2,\ldots,n$, with $P_t^{\rm(j)}$ changing between slots.  The sequence of variances employed across the slots, $P^{\rm (j)}_t,~t=-\frac{T}{2}, -\frac{T}{2}+1, \ldots, -1, 0, 1, \ldots, \frac{T}{2}-2, {\frac{T}{2} -1} $ is an i.i.d. sequence of uniform random variables on $[0, P_{\rm{max}}]$, where $P_{\rm{max}}$, as defined in Section II, is the maximum average power per symbol that the jammer can employ.   

Per above, Alice's codebook is only shared with Bob and thus is unknown to Willie.  However, Willie knows everything else about how the system is constructed, including the length of the codeword $n$, the distribution from which the codeword symbols are drawn (including $P_{\rm f}$), the distribution of the jamming power (including $P_{\rm{max}})$, the time of Alice's potential transmission, and his distances from Alice and the jammer.  Next, we establish that Willie's optimal strategy for detecting Alice's transmission is a power detector.



\begin{lem} \label{lem:opt_det_awgn}
{ Under assumptions of the AWGN model and the construction given above, }Willie's optimal detector compares the total received power in slot $t=0$ to a threshold.
\end{lem}


\noindent
\textbf{\emph{Proof: }}
Consider Willie's attempt to detect Alice during the slot $t=0$ of interest.  Since the jammer's power outside of this slot is independent of the jammer's power within the slot and since Willie knows $\sigma_{\rm w}^2$, it is sufficient for Willie to consider the vector of observations {$\bZ_0$} only within slot $t=0$, as defined in Section II.  
Hence, to simplify notation, we drop the slot index and denote the input to Willie's receiver as {$\bZ= [Z_1, Z_2, \ldots, Z_n]$}.

Given the assumptions of the lemma, the distribution of {$\bZ$} is complex Gaussian.  Under $H_0$, Willie observes only the jamming signal in addition to background noise.  Under $H_1$, Willie observes both the jamming signal and Alice's transmission in addition to background noise.   Let $\theta$ denote the variance of the power observed due to Alice's transmissions and the jammer's signal and thus define {$\bZ (\theta )= [Z_1(\theta ), Z_2(\theta ), \ldots, Z_n(\theta )]$, where $Z_i(\theta )\sim {\cal C} {\cal N}(0,\sigma_{\rm w}^2+\theta )$}. We thus distinguish between $H_0$ and $H_1$ by introducing two non-negative valued random variables $\Theta_0$ and $\Theta_1$ with probability density functions:
\begin{equation} \label{eq:awgnpdf}
f_{\Theta_\rho} (\theta ) = 
\begin{cases}
1/\zeta, & 0 < \theta \le P_{\rm{max}}/d^{\alpha}_{\rm j,w}, \rho=0\\
 1/\zeta, & \sigma_{\rm a}^2 < \theta \le \sigma^2_{\rm a}  + P_{\rm{max}}/d^{\alpha}_{\rm j,w}  ,\rho=1, \\
0, & \text{otherwise},
\end{cases}
\end{equation}
where $\zeta = P_{\rm{max}}/d_{\rm j,w}^{\alpha}$  and $\sigma_{\rm a}^2 =P_{\rm f} / d^{\alpha}_{\rm a,w}$.  The pdf of Willie's observations conditioned on $\theta$ is:
\begin{align}
f_{\bZ( \theta )}(\bf z)  & = \prod_{i=1}^n\frac{1}{{\pi (\sigma^2_{\rm w} + \theta)}}\exp\left(-\frac{|z_i|^2}{(\sigma^2_{\rm w} + \theta)} \right)\nonumber \\
& = \left(\frac{1}{\pi (\sigma^2_{\rm w} + \theta)}\right)^{n}\exp\left(-\frac{z}{(\sigma^2_{\rm w} + \theta)} \right),
\end{align}
where $z = \sum_{i=1}^n |z_i|^2$.  Thus, by the Fisher-Neyman Factorization Theorem, the total power $Z (\theta )= \sum_{i=1}^n |Z_i(\theta )|^2$ is a sufficient statistic for Willie's test. Let   $\chi^2_l$ denote a chi-squared random variable with $l$ degrees of freedom. Then $Z(\theta )= (\sigma_{\rm w}^2+\theta)\chi^2_{2n}$.  
Since Willie does not know either $\Theta_0$ or $\Theta_1$, his LRT becomes:
\[
\Lambda (Z) = \frac{E_{\Theta_1} [f_{Z(\theta )}(Z)]}{E_{\Theta_0} [f_{Z(\theta )}(Z)]}
\mathop{\gtrless}_{H_0}^{H_1} \gamma  .
\]
Next we show that $\Lambda (\cdot)$ is monotone.    From the definition of a chi-squared random variable, $Z(\theta ) \le_{\rm lr} Z(\theta ')$ whenever $\theta \le \theta '$.  In addition, applying the definition of $\le_{\rm lr}$ to the densities of $\Theta_0,\Theta_1$ yields that $\Theta_0 \le_{\rm lr} \Theta_1$.  The application of Lemma \ref{lem:lrmixtures} then yields that $\Lambda (\cdot)$ is non-decreasing in $z$. Thus, the LRT is equivalent to the test:
\[ Z \mathop{\gtrless}_{H_0}^{H_1} \Gamma_n \]
corresponding to a threshold test on the total received power.
\done

\begin{thm}\label{thm:awgn_covert}
Under the assumptions of the AWGN model, there exists a communication strategy for Alice, Bob, and the jammer whereby Alice transmits $\mathcal{O}(n)$ bits in $n$ channel uses reliably and covertly to Bob in the presence of Willie.
\end{thm}

\noindent
\textbf{\textit{Proof:}}
{\em Construction:}  Alice and the jammer employ the construction given at the beginning of Section III.  Per Lemma \ref{lem:opt_det_awgn}, the optimal detector for Willie is to employ a threshold test $Z \mathop{\gtrless}_{H_0}^{H_1} \Gamma_n$ on the total received power.  Dividing both sides by $n$ yields the equivalent test:
\begin{align}
\frac{Z}{n}~ \mathop{\gtrless}_{H_0}^{H_1} \tau_n,
\end{align}
where $\tau_n \equiv \Gamma_n / n$.  Whereas there is an optimal $\tau_n$ for any finite $n$, we will establish for any sequence of $\tau_n$ that Willie chooses, the detector is asymptotically useless as $n \rightarrow \infty$; that is, for any $\epsilon > 0$, there exists a construction such that $\mathbb{P}_\FA + \mathbb{P}_\MD > 1 - \epsilon$ for sufficiently large $n$.

{\em Analysis:}  Note that $\sigma_{\rm j}^2 = U \zeta$, where $U$ is a uniform random variable on $[0,1]$.  Recall that Willie does not know the value of $U$.   Let $\Prob_\FA (u)$ and $\Prob_{\rm MD}(u)$ be Willie's probability of false alarm and probability of missed detection conditioned on $U=u$, respectively.  Then,
\begin{align}
\mathbb{P}_{\rm FA}(u) = P\left (\frac{Z}{n} \geq \tau_n | U=u, H_0 \right).
\end{align}
Recall that $\chi^2_l$ denotes a chi-squared random variable with $l$ degrees of freedom.  Under $H_0$ and given $U=u$, $Z=(\sigma_{\rm w}^2 + u \zeta) \chi^2_{2n}$ and $Z/n=(\sigma_{\rm w}^2 + u \zeta) \chi^2_{2n}/n$.
By the weak law of large numbers, $\chi^2_{2n}/n$ converges in probability to 1; hence, for any $\delta >0$, $\exists N_0$ (not dependent on $u$) such that, for $n \geq N_0$,
\begin{align}
P \left (\frac{\chi^2_{2n}}{n} \in \left (1 - \frac{\delta}{\sigma_{\rm w}^2 + \zeta}, 1 + \frac{\delta}{\sigma_{\rm w}^2 + \zeta} \right ) \right) > 1 - \frac{\epsilon}{2}. 
\end{align}
Hence, for any $n > N_0$,
\begin{align}
P &\Bigg(\frac{Z}{n} \in \bigg((\sigma_{\rm w}^2 + u \zeta)\bigg (1 - \frac{\delta}{\sigma_{\rm w}^2 + \zeta}\bigg), \nonumber \\
&\qquad\qquad (\sigma_{\rm w}^2 + u \zeta) \bigg(1 + \frac{\delta}{\sigma_{\rm w}^2 + \zeta}\bigg) \bigg) \Bigg) > 1 - \frac{\epsilon}{2}.
\end{align}

\noindent {Since $u \leq 1$, $\sigma_{\rm w}^2 + u \zeta < \sigma_{\rm w}^2 + \zeta$ and thus,}
\begin{align}
P \left (\frac{Z}{n} \in \left (\sigma_{\rm w}^2 + u \zeta - \delta, \sigma_{\rm w}^2 + u \zeta + \delta \right ) \right ) > 1 - \frac{\epsilon}{2}.
\end{align}
Therefore, $\mathbb{P}_{\rm FA}(u) \geq 1 - \epsilon/2$ for any $\tau_n < \sigma_{\mathrm{w}}^2 + u \zeta - \delta$.  Likewise, following analogous arguments, there exists $N_1$ such that, for any $n > N_1$ (not dependent on $u$):
\begin{align}
\mathbb{P}_{\rm MD}(u) = P\left(\frac{Z}{n} \leq \tau_n | U=u, H_1 \right) > 1 - \frac{\epsilon}{2}
\end{align}
for any $\tau_n > \sigma_{\rm w}^2 + u \zeta + \sigma_{\rm a}^2 + \delta$.  Define the set $\mathcal{A}=\{u: \sigma_{\rm w}^2 + u \zeta - \delta < \tau_n < \sigma_{\rm w}^2 + u \zeta + \sigma_{\rm a}^2 + \delta\}$.  We have established that, for any $u\in \mathcal{A}^c$ and any $n > \mbox{max}(N_0,N_1)$, $\mathbb{P}_{\rm FA}({u}) + \mathbb{P}_{\rm MD}({u}) \geq 1 - \frac{\epsilon}{2}$.
The probability of {event $\mathcal{A}$} is bounded as:
\begin{align}
P(\mathcal{A}) & = P \left (\frac{\tau_n - \sigma_{\rm w}^2 - \sigma_{\rm a}^2 - \delta}{\zeta} \leq U \leq \frac{\tau_n - \sigma_{\rm w}^2 + \delta}{\zeta} \right ) \nonumber \\
       & \leq \frac{\sigma_{\rm a}^2 + 2 \delta}{\zeta}.
\end{align}
Hence, choosing $\delta = \zeta\epsilon/8$ and $\sigma_{\rm a}^2 = \zeta\epsilon/4$ yields:
\begin{align}
P({ \mathcal{A}^c}) \geq 1 - \frac{\epsilon}{2}. \label{eq:p_a_bar}
\end{align}
Therefore, the summation of Willie's false alarm and missed detection is lower-bounded as:
\begin{align}
\mathbb{P}_{\rm FA} + \mathbb{P}_{\rm MD} & = E_U\left [\mathbb{P}_{\rm FA}(U) + \mathbb{P}_{\rm MD}(U) \right] \\
                                                   & \geq E_U \left [\mathbb{P}_{\rm FA}(U) + \mathbb{P}_{\rm MD}(U)|{\mathcal{A}^c} \right] P({ \mathcal{A}^c})  \label{eq:pfa_pmd_1}\\ 
                                                   & > 1 - \epsilon.
\end{align}
Hence, Alice can employ codebooks with power $P_{\rm f} = \sigma_{\rm a}^2 d^{\alpha}_{\rm a,w}$ and remain covert from Willie.  Recognizing that the maximum interference caused by the jammer at Bob can be upper-bounded and hence the received signal-to-noise ratio at Bob can be lower-bounded by a constant, Alice can transmit $\mathcal{O}(n)$ bits in $n$ channel uses covertly and reliably to Bob. \done

\section{$M=1$ Block Fading Channels}\label{sec:block1}
\label{sec:blockm1}

\subsection{Covertness with Transmit Power not Decreasing in the Blocklength}

Recall that there are four channels in the problem formulation:  Alice-to-Bob, Alice-to-Willie, jammer-to-Bob, and jammer-to-Willie.  In this section, we expand the channel model to consider the situation where one or more of the four channels is a fading channel.   As in Section \ref{sec:awgn}, the problem is investigated by first characterizing how the Alice-to-Willie and jammer-to-Willie channels constrain (or not) the allowable scheme at Alice, in particular the power that she is able to employ while remaining covert.  The achievable performance under various metrics when Alice employs that power then follows classical information and communication theory based on the nature of the Alice-to-Bob and jammer-to-Bob channels.

Consider first the case where the Alice-to-Willie channel is an AWGN channel and the jammer-to-Willie channel is a $M=1$ block fading channel.  From an application perspective, this appears at first to be a pessimistic case:  the jammer who Alice is counting on to confuse Willie is subject to fading, whereas Willie has a strong direct path from Alice that makes the Alice-to-Willie channel comparatively benign (AWGN).  As in the case when all of the channels are AWGN, we first demonstrate that the optimal receiver at Willie is a power detector.  Unlike in Section \ref{sec:awgn}, here the jammer can transmit Gaussian noise drawn from a distribution with constant variance { $P_{\rm j}=P_{\rm{max}}$}, since the channel randomizes the power received at Willie from the jammer.

\begin{lem} \label{lem:opt_det_M=1}
Under the assumptions of the $M=1$ block fading model and Alice's construction presented in Section \ref{sec:awgn} but with the jammer transmitting Gaussian noise drawn from a distribution with constant variance, Willie's optimal detector for detecting Alice's transmission is to compare the total received power in the slot of interest to a threshold.
\end{lem}

\noindent
\textbf{\textit{Proof:}}
Let $\zeta = P_{\rm j} / d_{\rm j,w}^{\alpha}$.   The received jammer power $\sigma_{\rm j}^2$ is exponentially distributed with mean $\zeta$.  As in Section III, note that observations outside of $k=1,2,\ldots,n$ do not help Willie to detect a transmission by Alice in slot $t=0$; hence, it is sufficient to consider $\bZ_0$ as the input to Willie's receiver.  We therefore suppress the slot index and denote Willie's observation conditioned on $\theta$ by $\bZ(\theta) = [Z_1(\theta), Z_2(\theta), \ldots, Z_n(\theta)]$ where $Z_i(\theta )\sim {\cal C} {\cal N}(0,\sigma_{\rm w}^2+\theta )$.   We distinguish between $H_0$ and $H_1$ by introducing two non-negative valued random variables $\Theta_0$ and $\Theta_1$ with probability density functions:
\begin{equation} \label{eq:M=1pdf}
f_{\Theta_\rho} (\theta ) = 
\begin{cases}
\frac{1}{\zeta} e^{-\theta /\zeta}, & 0 < \theta, \rho=0,\\
\frac{1}{\zeta} e^{-(\theta - \sigma^2_{\rm a})/\zeta}, & \sigma_{\rm a}^2 < \theta ,\rho=1, \\
0, & \text{otherwise}.
\end{cases}
\end{equation}

{ Thus, $\Theta_0\le_{\rm lr} \Theta_1$ based on the assumptions presented in Section II.  The distribution of Willie's observations conditioned on $\theta$ is:   }  
\begin{equation}
f_{\bZ(\theta)}(\mathbf{z}) = \Bigl(\frac{1}{\pi(\sigma^2_{\rm w} + \theta )}\Bigr)^n\exp\Bigl(-\frac{z}{\sigma^2_{\rm w} + \theta } \Bigr),
\end{equation}
where $z$ is as defined in Section \ref{sec:awgn}.  Hence, the LRT test is optimal based on the NP rule and the optimal decision rule for Willie again becomes:
\begin{equation} \label{LR:M=1}
\Lambda (Z) = \frac{E_{\Theta_1} [f_{Z(\theta )}(Z)]}{E_{\Theta_0} [f_{Z(\theta )}(Z)]}
\mathop{\gtrless}_{H_0}^{H_1} \gamma  .
\end{equation}
The monotonicity of $\Lambda (\cdot)$ then follows from Lemma \ref{lem:lrmixtures} by observing that, as in the proof of Lemma 2, $Z(\theta ) \le_{\rm lr} Z(\theta ')$ whenever $\theta \le \theta '$, and, as noted above, $\Theta_0 \le_{\rm lr} \Theta_1$.  Thus, the LRT is equivalent to the power detector: $Z \mathop{\gtrless}_{H_0}^{H_1} \Gamma_n$.
\done

Next, we consider the case when the Alice-to-Willie channel is also a $M=1$ block fading channel.  In practice, Willie does not know the value of the fading coefficient $h_{0,1}^{\rm (a,w)}$ on this channel and, indeed, that is our assumption in our achievability result below.   However, since we are interested in an achievability result for covert communication from Alice to Bob, giving Willie any extra knowledge (say, by a genie) only strengthens the result.  Hence, in the Corollary below, which we use below to establish Theorem 2, we assume Willie knows $h_{0,1}^{\rm (a,w)}$.

\begin{cor} \label{cor:opt_det_M=1_all_fades}
Consider the assumptions of the model when the jammer-to-Willie and Alice-to-Willie channels are block fading channels with one fading block per codeword.  Additionally, assume that Willie knows the value of $h_{0,1}^{\rm (a,w)}$.  Then, given Alice's construction in Section \ref{sec:awgn} but with the jammer transmitting Gaussian noise drawn from a distribution with constant variance, Willie's optimal detector for detecting a transmission by Alice is to compare the total received power in the slot of interest to a threshold.
\end{cor}

\noindent\textbf{\textit{Proof:}} Knowing $h^{\rm (a,w)}_{0,1}$ and $d_{\rm a,w}$, Willie knows $\sigma_{\rm a}^2$, and the proof follows from Lemma \ref{lem:opt_det_M=1}.
\done
\begin{thm} \label{thm:covert_M=1}
Under the assumptions of the single block fading model, there exists a communication strategy for Alice, Bob, and the jammer whereby Alice transmits with a power that does not decrease with the blocklength while remaining covert from warden Willie. 
\end{thm}

\noindent\textbf{\textit{Proof:}}  This proof follows along the lines of Theorem 1 and is provided in Appendix A.\done


\subsection{The Number of Covert Bits Transmitted Reliably}\label{sec:covert_thru}

Theorem 2 establishes that Alice can transmit with power not decreasing in the blocklength $n$ while maintaining covertness.  In the case of AWGN channels on both the Alice-to-Bob and jammer-to-Bob channels, the covert and reliable communication of $\mathcal{O}(n)$ bits in $n$ channel uses can be achieved.   However, when the Alice-to-Bob or jammer-to-Bob channels are $M$-block fading channels, $M \geq 1$, the problem is analogous to the standard problem of communication over slowly fading channels \cite[Section 5.4]{tse_vishwanath}.  Strictly speaking, reliable communication as defined in Section \ref{sec:metrics} of $\mathcal{O}(n)$ bits is not possible.  In particular, if Alice transmits $n R_0$ bits for any given constant $R_0 >0$, there always exists some nonzero probability, not diminishing in $n$, that the instantiations of $|h_{0,m}^{\rm (a,b)}|$ and $|h_{0,m}^{\rm (j,b)}|$, $m=1,2, \ldots M$, will lead to a received signal-to-interference-plus-noise ratio (SINR) such that the communication is not reliable.  

However, the presence of the jammer, which allows Alice to transmit at per-symbol power $P_{\rm f} >0$ not dependent on $n$ (versus $\mathcal{O}(\frac{1}{\sqrt n})$ power per symbol when there is no jammer\cite{bash_jsac}),  greatly improves system performance even in the case when the Alice-to-Bob or jammer-to-Bob channels are $M$-block fading channels.  This can be seen via multiple metrics.  First, if the metric of Section \ref{sec:metrics} is still of pertinent interest, covert and reliable communication of $o(n)$ bits is possible, as demonstrated for $M=1$ in Appendix B.  Second, and probably of more interest, is that the analog of the $\epsilon$-outage capacity (see \cite{tse_vishwanath}) is non-zero, whereas it would be zero for any transmission power at Alice that decreases to 0 as $n \rightarrow \infty$.   


\section{$M > 1$ Block Fading Channels}\label{sec:block2}
\label{sec:blockm2}

Here we consider the case of an $M > 1$ block fading channel on the jammer-to-Willie link.   In contrast to the results of Lemma \ref{lem:opt_det_awgn} and Lemma \ref{lem:opt_det_M=1} for the AWGN and $M=1$ block fading channels on the Alice-to-Willie link, respectively, a power detector is not the optimal detector for Willie.  Instead, we establish an important property of the optimal detector in Lemma \ref{lem:det_M>1}:  that, if a given vector of observed powers for the $M$ blocks encompassing a slot results in a point on the boundary between Willie's decision regions, an increase in any component of that vector results in a decision of $H_1$.  Whereas this does not explicitly identify the optimal receiver, it does guarantee an important property of the dividing ``curve'' between the two decision regions:  for any given $M-1$ components of the vector of observed powers, there is at most one solution for the remaining component that falls on this curve between $H_0$ and $H_1$, as defined precisely below.  In particular, this is then sufficient to establish the result of interest:  that Alice can transmit covertly at power that does not decrease with the blocklength $n$.


\subsection{Properties of the Optimal Detector at Willie}
With $t=0$ the slot of interest, observations outside of $k=1,2,\ldots,n$ do not help Willie detect transmissions by Alice in slot $t=0$.  Therefore, the slot index is suppressed, and we denote Willie's observations by $\mathbf{\hat Z} =  [\hat{Z}_1, \hat{Z}_2,\ldots,\hat{Z}_n]$.   Conditioned on the fading coefficients on the jammer-to-Willie channel, measurements within each fading block of length $n/M$ are i.i.d., but the measurements from different blocks come from different distributions determined by the sequence of block fading variables.  Therefore, when Alice does not transmit, Willie's observations have the distribution:
\begin{subequations}
\begin{align}
f_{\mathbf{\hat Z}|H_0}(\mathbf{\hat z}|H_0) &= E_{\mathbf{h}^{\rm(j,w)}} \Bigg[\prod_{m=1}^M \prod_{i=1}^{n/M} \frac{1}{\pi(\sigma_{\rm w}^2 + \sigma_{{\rm j},m}^2)} \nonumber\\
&\qquad\qquad\qquad\qquad \cdot e^{-\frac{|{\hat z}_{(m-1)\frac{n}{M}+i}|^2}{(\sigma_{\rm w}^2+\sigma_{{\rm j},m}^2)}} \Bigg] \label{eq:c1}\\
&= \prod_{m=1}^M E_{{h}^{\rm(j,w)}_m} \Bigg [ \left(\frac{1}{\pi (\sigma_{\rm w}^2 + \sigma_{{\rm j},m}^2)}\right)^{\frac{n}{M}} \nonumber\\
& \qquad\qquad\qquad\qquad\quad \cdot e^{-\frac{{z}_{m}}{(\sigma_{\rm w}^2+\sigma_{{\rm j},m}^2)}} \Bigg ], \label{eq:c2}
\end{align}
\end{subequations}
where $\mathbf{h}^{\rm(j,w)}= [h^{\rm(j,w)}_1,h^{\rm(j,w)}_2,\ldots,h^{\rm(j,w)}_M]$ is the vector of (complex) fading coefficients on the jammer-to-Willie channel, $z_{m} = \sum_{i=1}^{n/M}|{\hat z}_{(m-1)\frac{n}{M}+i}|^2$, and $\sigma_{{\rm j},m}^2 = \frac{P_{\rm j}^{(t)} |h^{\rm(j,w)}_m|^2}{d_{\rm j,w}^{\alpha}}$.  Let $\zeta = P_{\rm j}^{(t)} / d_{\rm j,w}^{\alpha}$ and $\mathbf{Z} = [Z_1, Z_2,\ldots,Z_M]$, where $Z_m=\sum_{i=1}^{n/M}|{\hat Z}_{(m-1)\frac{n}{M}+i}|^2$ is the power measured in the $m^{\rm th}$ block.  The distribution of the vector $\bZ$ of received powers across the $M$ blocks under $H_0$ is:
\begin{align}
f_{\mathbf{Z}|H_0}(\mathbf{z}|H_0) &= \frac{1}{\pi^n}\prod_{m=1}^M  \int_{0}^{\infty}\left(\frac{1}{\sigma_{\rm w}^2+u}\right)^{\frac{n}{M}} \nonumber\\
&\qquad\qquad\qquad\qquad\quad \cdot e^{ -\frac{z_m}{(\sigma_{\rm w}^2+u)}} e^{ -\frac{u}{\zeta}} du\\
&= \frac{e^{\frac{M \sigma_{\rm w}^2}{\zeta}}}{\pi^n}
\prod_{m=1}^M  \int_{\sigma_{\rm w}^2}^{\infty}\left(\frac{1}{v}\right)^{\frac{n}{M}}e^{ -\frac{z_m}{v}} e^{ -\frac{v}{\zeta}}dv.\label{eq:h0_gr1}
\end{align}
Similarly, the distribution under $H_1$ is:
\begin{align}
f_{\mathbf{Z}|H_1}(\mathbf{z}|H_1) &= \frac{e^{\frac{M (\sigma_{\rm w}^2+\sigma_{\rm a}^2)}{\zeta}}}{\pi^n}
 \prod_{m=1}^M  \int_{\sigma_{\rm w}^2+\sigma_{\rm a}^2}^{\infty}\left(\frac{1}{v}\right)^{\frac{n}{M}} \nonumber\\
 &\qquad\qquad\qquad\qquad\qquad \cdot e^{ -\frac{z_m}{v}} e^{ -\frac{v}{\zeta}}dv.\label{eq:h1_gr1}
\end{align}
The LRT test is then:
\begin{align}
\Lambda(\mathbf{Z}) = \frac{e^{\frac{M\sigma_{\rm a}^2}{\zeta}} \prod_{m=1}^M  \int_{\sigma_{\rm w}^2+\sigma_{\rm a}^2}^{\infty}\left(\frac{1}{v}\right)^{\frac{n}{M}} e^{-\frac{Z_m}{v}}e^{-\frac{v}{\zeta}}dv}{\prod_{m=1}^M  \int_{\sigma_{\rm w}^2}^{\infty}\left(\frac{1}{v}\right)^{\frac{n}{M}} e^{-\frac{Z_m}{v}}e^{-\frac{v}{\zeta}}dv} \mathop{\gtrless}_{H_0}^{H_1}\gamma. \label{LRT:M>1} 
\end{align}
The LRT in (\ref{LRT:M>1}) shows that $\bZ$ forms a sufficient statistic for the optimal test for Willie to determine whether Alice transmits in that slot or not.  The following lemma establishes that $\Lambda (\cdot)$ is monotone increasing in each of its components.




\begin{lem} \label{lem:det_M>1}
Consider the assumptions of the multiple block fading channel model and Alice's construction presented in Section \ref{sec:awgn}  but with the jammer transmitting Gaussian noise drawn from a distribution with constant variance.    When the Alice-to-Willie channel is AWGN and the jammer-to-Willie channel is faded,  $\Lambda (\bZ )$ is monotonically increasing in each of the components of $\bZ$.
\end{lem}

\noindent
\textbf{\textit{Proof:}} $\Lambda(Z)$ (defined in (\ref{LR:M=1})) monotonically increases in $Z$ in the $M=1$ case as shown in Appendix C.  The proof then follows from the observation that $\Lambda (\bZ )$ in the $M>1$ case can be expressed as:
\begin{align}
\Lambda (\bZ ) = \prod_{i=1}^M \Lambda(Z_i). \qquad\qquad \done \label{eq:lem4_test}
\end{align}

\begin{cor}\label{cor:det_M>1}
{ Consider the assumptions of the multiple block fading model and Alice's construction presented in Section \ref{sec:awgn} but with the jammer transmitting Gaussian noise drawn from a distribution with constant variance.  Additionally, assume that Willie knows $h_{0,m}^{\rm (a,w)}, m=1,2,\ldots,M$.  When fading exists on both the jammer-to-Willie channel and the Alice-to-Willie channel, then} the likelihood ratio $\Lambda (\bZ )$ is monotonically increasing in each of the components of $\bZ$.
\end{cor}

\textbf{\textit{Proof:}}  Conditioned on Willie's knowledge of $h_{0,m}^{\rm (a,w)}, m=1,2,\ldots,M$, the channel from Alice-to-Willie is an AWGN channel with a different signal power for Alice per block; hence, the result follows similarly to that of Lemma \ref{lem:det_M>1}.
\done

\subsection{Covertness with Transmit Power not Decreasing in the Blocklength}

Next, we leverage Lemma \ref{lem:det_M>1} on the structure of the optimal receiver at Willie to demonstrate the ability for Alice to employ power not decreasing in the blocklength for the case where there exists $M > 1$ block fading on the jammer-to-Willie channel.  The general concept of the proof is similar to Theorem 1:  demonstrate that the optimal detector at Willie works poorly on a set of fading instantiations of the jammer's signal that has high probability.

Before we outline the proof, we first need to define a number of regions that characterize Willie's detector.  Recall that a sufficient statistic for Willie's optimal detector is given by $\bZ = [Z_1,Z_2,\ldots ,Z_M]$, where $Z_i$ is the power measured in the $i^{\rm th}$ block.  A normalized version corresponding to the average observed power per symbol within a block is also a sufficient statistic for the optimal detector: $\bX = [X_1, X_2,\ldots ,X_M]$, where $X_{i} = \frac{Z_{i}}{n/M}, i=1,2,\ldots M$.  A detector for Willie is defined by the regions $R_{H_0}(n)$ and $R_{H_1}(n)$, each in ${\cal R}^M$, where $H_0$ is chosen if $\mathbf{X} \in R_{H_0}(n)$, and $H_1$ is chosen if $\mathbf{X} \in R_{H_1}(n)$.   For the optimal detector at Willie, as given in (\ref{LRT:M>1}), a vector $\bx$ is in $R_{H_1}(n)$ if and only if $\Lambda(\frac{n}{M}  \bx ) > \gamma$; otherwise ${\mathbf x}$ is in $R_{H_0}(n)$.  Hence, define the boundary curve dividing $R_{H_0}(n)$ and $R_{H_1}(n)$ as ${\rm C}(n) = \{\bx: \Lambda(\frac{n}{M}  \bx ) = \gamma$\}.  Finally, we define a boundary region, $R^{\delta}_{\rm B}(n)$, as the set of all $\bx$ that are within distance $\delta$ in each dimension of $C(n)$; that is:
\begin{align}
R^{\delta}_{\rm B}(n) = \{\bx: \exists~\by \in C(n) \mbox{ s.t. } \max_i |x_i-y_i| < \delta \}.
\end{align}


Define the $M$-dimensional vectors $\boldsymbol{\sigma}_{\rm j}^2 = [\sigma_{{\rm j},1}^2,\sigma_{{\rm j},2}^2, \ldots ,\sigma_{{\rm j},M}^2]$ and $\boldsymbol{\sigma}_{\rm w}^2 = \sigma_{\rm w}^2  [1,1,\ldots ,1]$.  Note that $\boldsymbol{\sigma}_{\rm j}^2$ is random, since it depends on the fading from the jammer to Willie, whereas $\boldsymbol{\sigma}_{\rm w}^2$ is deterministic and known to Willie.   The proof then proceeds, as follows.  Given the instantiation of the block fading values between the jammer and Willie, which determines the expected jammer power per symbol $\sigma^2_{{\rm j},i}$ for the $i^{\rm th}$ fading block, the $i^{\rm th}$ element of the vector $\mathbf{X}$ has the expected value $\sigma^2_{{\rm j},i}+ \sigma_{\rm w}^2$ (under $H_0$) or $\sigma^2_{{\rm j},i} +\sigma_{\rm w}^2 + \sigma_{\rm a}^2$ (under $H_1$).    The proof then begins with Lemma \ref{lem:M>1vanish}, which leverages Lemma \ref{lem:det_M>1} to show that the probability of fading instantiations that result in $\boldsymbol{\sigma}^2_{\rm j} +\boldsymbol{\sigma}^2_{\rm w}  \in R^{\delta}_{\rm B}(n)$ can be made arbitrarily small by choosing $\delta$ small enough; hence, the probability that the jamming is such that the average power received per symbol when Alice is not transmitting is in the boundary region can be made arbitrarily small.  The theorem then follows by considering what happens for the (highly probable) event that the instantiation of the block fading values yields $\boldsymbol{\sigma}^2_{\rm j} +\boldsymbol{\sigma}^2_{\rm w}  \notin R^{\delta}_{\rm B}(n)$; in this case, for $\sigma_{\rm a}^2$ sufficiently small, the probability of missed detection or the probability of false alarm is near one.  Hence, Alice can employ power {that does not decrease with $n$} and still achieve covertness.  Essentially, Willie is not able to set a boundary curve that works for a large set of $\boldsymbol{\sigma}_{\rm j}^2$, and thus his detector is only effective in the unlikely event that $\boldsymbol{\sigma}^2_{\rm j} +\boldsymbol{\sigma}^2_{\rm w}$ is near the boundary curve between his decision regions.

\begin{lem} \label{lem:M>1vanish} 
 Under the assumptions of the multiple block fading model,   for Willie's optimal detector, with $R^{\delta}_{\rm B}(n)$ as defined above, for any $\epsilon > 0$ there exists $\delta > 0$ s.t. $P(\mathbf{h}: \boldsymbol{\sigma}^2_{\rm j} +\boldsymbol{\sigma}^2_{\rm w} \in R^{\delta}_{\rm B}(n)) < \epsilon$.
\end{lem}

\noindent
\textbf{\textit{Proof:}}
See Appendix D.  \done

\begin{thm}\label{thm:covert_M>1}
Consider the assumptions of the multiple block fading model and Alice's construction in Section \ref{sec:awgn} but with the jammer transmitting Gaussian noise drawn from a distribution with constant variance.   Then, there exists a communication strategy for Alice, Bob, and the jammer whereby Alice transmits with a power that does not decrease with the blocklength while being covert from Willie.
\end{thm}

\noindent
\textbf{{\em Proof:}} 
Consider a covertness criterion $\mathbb{P}_{\rm MD} + \mathbb{P}_{\rm FA} > 1 - \epsilon$.   
By Lemma \ref{lem:M>1vanish}, choose $\delta > 0$ s.t.:
\begin{align}
P(\mathbf{h}: \boldsymbol{\sigma}^2_{\rm j} +\boldsymbol{\sigma}^2_{\rm w} \in R^{2\delta}_{\rm B}(n)) < \frac{\epsilon}{4}.
\end{align}

If the Alice-to-Willie channel is AWGN, choose constant $P_{\rm f} > 0$ such that $\sigma_{\rm a}^2 < \delta$.  If the Alice-to-Willie channel is a $M \geq 1$ block fading channel, choose $P_{\rm f} > 0$ such that the average received power from Alice is less than $\delta$ for all fading blocks with high probability.   We proceed with the proof for the case when the Alice-to-Willie channel is AWGN, but the modifications for when the Alice-to-Willie channel is a $M \geq 1$ block fading channel  follow similar steps to those shown in the second part of the proof of Theorem 2 in Appendix A.

Consider an optimal detector at Willie for blocklength $n$, with associated decision regions $R_{H_0}(n)$ and $R_{H_1}(n)$.  First, we present a sketch of the proof idea.  Consider the case where $\boldsymbol{\sigma}_{\rm w}^2 + \boldsymbol{\sigma}_{\rm j}^2 \in R_{H_0}(n) \setminus R^{2 \delta}_{\rm B}(n) $.  If Alice is employing $\sigma_{\rm a}^2 < \delta$, the probability of Willie's test result being in $R_{H_1}(n)$ occurs with small probability for large $n$, regardless of whether $H_0$ or $H_1$ is true.  Thus, Willie's $\mathbb{P}_{\rm MD}$ will be large and $\mathbb{P}_{\rm FA}$ will be small.  Likewise, if $\boldsymbol{\sigma}_{\rm w}^2 + \boldsymbol{\sigma}_{\rm j}^2 \in R_{H_1}(n) \setminus R^{2 \delta}_{\rm B}(n)$, then Willie's $\mathbb{P}_{\rm FA}$ will be large and $\mathbb{P}_{\rm MD}$ will be small for large $n$.  

The rigorous proof is the vector extension of that of Theorem 2.  Recall that $[\sigma_{\rm j,1}^2, \sigma_{\rm j,2}^2,\ldots, \sigma_{\rm j,M}^2]$ is an i.i.d. vector, where each component is exponentially distributed with mean $\zeta$.  Hence, there exists a constant $c$ s.t.
\begin{align}
P\left(\max_{i=1,2,\ldots,M} ~\sigma_{\rm j,i}^2 > c\right) < \frac{\epsilon}{4}.
\end{align} 
Let
\begin{align}
\mathbb {P}_{\rm FA}({\mathbf u}) = P({\mathbf X} \in R_{H_1}(n) | \boldsymbol{\sigma}_{\rm j}^2 + \boldsymbol{\sigma}_{\rm w}^2={\mathbf u}, H_0).
\end{align}
Under $H_0$, $X_i = (\sigma_{\rm w}^2 + \sigma_{\rm j,i}^2) \chi_{\frac{2n}{M}, i}^2, i=1,2,\ldots, M$, where $\{
\chi_{\frac{2n}{M}, i}^2,i=1,2,\ldots, M\}$ is an i.i.d.~collection of (central) chi-squared random variables, each with $2n/M$ degrees of freedom.  By the weak law of large numbers, each converges in probability to 1; since $M$ is finite, this implies $\exists N_0$ s.t. $\forall n \geq N_0$,
\begin{align}
P \left(\bigcap_{i=1}^M \left \{\chi_{\frac{2n}{M}, i}^2 \in \left (1 - \frac{\delta}{\sigma_{\rm w}^2 +c}, 1 + \frac{\delta}{\sigma_{\rm w}^2 +c}\right ) \right\}\right) > 1- \frac{\epsilon}{2},
\end{align}
and
\begin{align}
P &\Bigg(\bigcap_{i=1}^M \bigg\{X_i \in \bigg((\sigma_{\rm w}^2 + \sigma_{\rm j,i}^2) \left (1 - \frac{\delta}{\sigma_{\rm w}^2 +c} \right ), \nonumber \\
&\qquad\qquad(\sigma_{\rm w}^2 + \sigma_{\rm j,i}^2) \left (1 + \frac{\delta}{\sigma_{\rm w}^2 +c} \right)\bigg ) \bigg\}\Bigg) > 1- \frac{\epsilon}{2}.
\end{align}
Now, if $\max_{i=1,2,\ldots,M} ~\sigma_{\rm j,i}^2 \leq c$, then $\sigma_{\rm w}^2 + \sigma_{\rm j,i}^2 < \sigma_{\rm w}^2 +c$, and thus, for $n \geq N_0$:
\begin{align}
P \left(\bigcap_{i=1}^M \left \{X_i \in (\sigma_{\rm w}^2 + \sigma_{\rm j,i}^2 - \delta, \sigma_{\rm w}^2 + \sigma_{\rm j,i}^2 + \delta) \right\}\right) > 1- \frac{\epsilon}{2}.
\end{align}
Thus, if ${\mathbf u} \in R_{H_1} \setminus R^{2 \delta}_{\rm B}(n)$, then $P({\mathbf X} \in R_{H_1}) > 1 - \frac{\epsilon}{2}$ and
\begin{align}
\mathbb{P}_{\rm FA}({\mathbf u}) > 1 - \frac{\epsilon}{2}.
\end{align}
Next consider any ${\mathbf u} \in R_{H_0} \setminus R^{2 \delta}_{\rm B}(n)$.  Then, recalling $\sigma_{\rm a}^2 < \delta$,  the vector ${\mathbf u} + \sigma_{\rm a}^2 [1~ 1~ \ldots ~1]$ cannot have any element within $\delta$ of ${\rm C}(n)$.  Then, following analogous arguments to those above,  $\exists N_1$ s.t. for $n \geq N_1$,
\begin{align}
\mathbb{P}_{\rm MD}({\mathbf u}) & = P({\mathbf X} \in R_{H_0}(n) | \boldsymbol{\sigma}_{\rm j}^2 + \boldsymbol{\sigma}_{\rm w}^2={\mathbf u}, H_1) \\
& > 1 - \frac{\epsilon}{2}
\end{align}
for ${\mathbf u} \in R_{H_0} \setminus R^{2\delta}_{\rm B}(n)$ whenever $\max_{i=1,2,\ldots,M} ~\sigma_{\rm j,i}^2 \leq c$.
Thus, unless 
\begin{align}
{\mathcal A} = \left\{{\mathbf u} \in R^{2\delta}_{\rm B}(n)\right\} \cup \left\{\max_{i=1,2,\ldots,M} ~\sigma_{\rm j,i}^2 > c\right\}
\end{align}
occurs, 
\begin{align}
\mathbb{P}_{\rm FA}({\mathbf u}) + \mathbb{P}_{\rm MD}({\mathbf u}) > 1 - \frac{\epsilon}{2}.
\end{align}
By construction, $P({\mathcal A}) < \epsilon / 2$, and thus
\begin{align}
\mathbb{P}_{\rm FA} + \mathbb{P}_{\rm MD} & = E_U[\mathbb{P}_{\rm FA}({\mathbf U}) + \mathbb{P}_{\rm MD}({\mathbf U})] \\
& \geq E_U[\mathbb{P}_{\rm FA}({\mathbf U}) + \mathbb{P}_{\rm MD}({\mathbf U}) | {\mathcal A}^c] P({\mathcal A}^c) \\
& > 1 - \epsilon.
\end{align}
\done

The implications on reliable throughput are then analogous to those discussed in Section \ref{sec:covert_thru}.

\section{Discussion}

\label{sec:discussion}
\subsection{Active Adversary May Help Covert Communication}
The assumptions presented in Section \ref{sec:model} assume that the jammer is attempting to help Alice and Bob to communicate covertly.  However, covert communication may still be possible if an adversarial jammer is placed in the environment to actively try to jam any potential communication by Alice, as is commonly done in electronic warfare.  For example, suppose that Willie uses a jammer to inhibit communication by any party; then, whereas this jammer does indeed decrease the rate of any reliable (non-covert) communication, it may actually facilitate covert communication by hurting Willie's ability to determine if Alice is transmitting. In particular, if the jammer-to-Willie channel is faded and Willie's jammer transmits Gaussian noise, then exactly the same interference model as derived for the constructions of Sections \ref{sec:block1} and \ref{sec:block2} applies.  This enables covert communication from Alice to Bob in precisely the same manner as in the case of a ``friendly'' jammer. Note that this assumes that such a jammer generates random Gaussian noise; if that jammer instead generates a noise-like signal that is decodable by Willie (say, using a Gaussian codebook shared by the jammer and Willie), then Willie can conceivably decode the jammer's signal and subtract it from his received signal, subject only to the standard challenges of successive interference cancellation in wireless communication environments.

\subsection{Relationship with Steganography}
Steganography is the discipline of hiding messages in innocuous objects. Typical steganographic systems modify fixed-size finite-alphabet covertext objects into stegotext containing hidden information, and are subject to a similar square root law (SRL) as non-jammer assisted covert communication: $\mathcal{O}(\sqrt{n})$ symbols in size $n$ covertext may safely be altered to hide an $\mathcal{O}(\sqrt{n}\log n)$-bit message \cite{fridrich09stego}. As explained in \cite{bash_jsac}, the mathematics of statistical hypothesis testing are responsible for both SRLs while the extra $\log n$ factor is from the lack of noise in the steganographic context.  However, arguably the earliest work on SRL shows that it is achievable without the $\log n$ factor when an active adversary corrupts stegotext with AWGN \cite{korzhik05srl}.\footnote{We note that the results of \cite{bash_isit2012} and \cite{bash_jsac} were developed independently of \cite{korzhik05srl}. While \cite{korzhik05srl} provides the proof of the SRL when Alice is average-power constrained, \cite{bash_isit2012} and \cite{bash_jsac} also develop the achievability of SRL for the peak-power constained covert communication and the converse to the SRL.} That being said, \cite{craver10nosqrtlaw} shows that, because Alice in the steganographic setting has write-access to covertext, the SRL can be broken and $\mathcal{O}(n)$ bits can be embedded in size $n$ covertext using careful selection of the subset of the covertext to be overwritten \cite{craver10nosqrtlaw}. Thus, unlike the scenario considered here, breaking the steganographic SRL does not require Willie to be uncertain about the distribution of his observations. 

\section{Conclusion}
In this paper, we have considered the ability for Alice to transmit covertly and reliably to Bob with the help of a jammer in the presence of a watchful adversary Willie.  For either an AWGN or block fading channel between the jammer and Willie, under the assumption of a key of unlimited length shared between Alice and Bob, we are able to establish that Alice can transmit with power not decreasing in the blocklength $n$ while remaining covert, even when Willie employs an optimal receiver.   In the case of AWGN channels from Alice to Bob and the jammer to Bob, this implies positive rate covert communication.  In the case of fading channels on either the Alice-to-Bob link or the jammer-to-Bob link, standard communication results for communication over fading channels are achievable.

Whereas the wireless communication channel models presented here are standard practice for the design of reliable communication systems, their mapping to the covert communication problem motivates further study.  In particular, the assumption of block fading, which results in the jammer power outside of the codeword slot of interest being independent of that within the codeword slot of interest, needs to be carefully examined.  If the block fading model is too optimistic for covert communication, a potential solution would be for the jammer to randomly vary his/her power in each codeword slot as is done here in the AWGN case.   Hence, we feel the most important assumption to be relaxed in future work is that of synchronism between the slot boundaries at Alice and the jammer.   Whereas this assumption certainly seems reasonable given the accuracy of modern clocks, small errors might allow the adversary Willie to perform estimation of the environment that could inhibit covert communication, and thus, while complicating the model and requiring assumptions on current technology, this deserves careful consideration.  Future work will also investigate the achievable performance for covert communications under limitations on the size of the shared key between Alice and Bob.

\bibliographystyle{IEEEtran} 
\bibliography{covert}


\section*{Appendix}
\renewcommand{\thesubsection}{\Alph{subsection}}
\numberwithin{equation}{subsection}

\subsection{\bf Proof of Theorem 2}
{\em Construction:}  Alice and the jammer employ the same methods as described in the construction of Lemma 3.  Hence, Willie is aware that the channel gain between the jammer and himself results in $\sigma_{\rm j}^2$ being distributed as an exponential random variable with mean $\zeta$.  If the Alice-to-Willie channel is AWGN, Lemma 3 establishes that the optimal receiver for Willie to employ is a power detector  $Z \mathop{\gtrless}_{H_0}^{H_1} \Gamma_n$ for some threshold $\Gamma_n$ on the slot of size $n$, or, equivalently,
\begin{align}
\frac{Z}{n} \mathop{\gtrless}_{H_0}^{H_1} \tau_n,
\label{pow_detect_app}
\end{align}
where $\tau_n \equiv \Gamma_n/n$.
If the Alice-to-Willie channel is an $M=1$ block fading channel, we assume pessimistically that Willie also knows the value of $h_{0,1}^{\rm (a,w)}$.  Then, Corollary 3.1 establishes that the optimal receiver for Willie is again the power detector in 
(\ref{pow_detect_app}).

\noindent {\em Analysis}:
Consider first the case when the Alice-to-Willie channel is an AWGN channel.  Recall that we require $\mathbb{P}_{\rm FA} + \mathbb{P}_{\rm MD} > 1 - \epsilon$ for any $\epsilon > 0$.  Thus, consider any $\epsilon > 0$.  The unboundedness of the support of $\sigma_{\rm j}^2$ requires a slight modification of the proof technique of Theorem 1.  Thus, note that there exists some constant $c$ such that:
\begin{align}
P( \sigma_{\rm j}^2 > c) < \epsilon/4. \label{eq:c_gr_sigmaj}
\end{align}
Consider first the false alarm rate, and, analogously to the proof of Theorem 1, define:
\begin{align}
\mathbb{P}_{\rm FA}(u) = P\left (\frac{Z}{n} \geq \tau_n | \sigma_{\rm j}^2=u, H_0 \right).
\end{align}
Under $H_0$, $Z/n = (\sigma_{\rm w}^2+\sigma_{\rm j}^2)\chi^2_{2n}/n$.  By the weak law of large numbers, $\chi^2_{2n}/n$ converges in probability to 1; hence, for any $\delta >0$, $\exists N_0$ (not dependent on $u$) such that, for $n \geq N_0$,
\begin{align}
P \left (\frac{\chi^2_{2n}}{n} \in \left (1 - \frac{\delta}{\sigma_{\rm w}^2 + c}, 1 + \frac{\delta}{\sigma_{\rm w}^2 + c} \right ) \right) > 1 - \frac{\epsilon}{2}.
\end{align}
Hence, for any $n > N_0$,
\begin{align}
P &\Bigg (\frac{Z}{n} \in \bigg((\sigma_{\rm w}^2 + u)\left (1 - \frac{\delta}{\sigma_{\rm w}^2 + c}\right ), \nonumber\\
&\qquad\qquad(\sigma_{\rm w}^2 + u) \left (1 + \frac{\delta}{\sigma_{\rm w}^2 + c}\right )  \bigg) \Bigg) > 1 - \frac{\epsilon}{2}.
\end{align}
Now, for any $u \leq c$, $\sigma_{\rm w}^2 + u < \sigma_{\rm w}^2 + c$ and thus for any $n > N_0$:
\begin{align}
P \left (\frac{Z}{n} \in \left (\sigma_{\rm w}^2 + u - \delta, \sigma_{\rm w}^2 + u + \delta \right ) \right ) > 1 - \frac{\epsilon}{2}
\end{align}
and thus $\mathbb{P}_{\rm FA}(u) \geq 1 - \epsilon/2$ for any $\tau_n < \sigma_{\mathrm{w}}^2 + u  - \delta$ as long as $u < c$.
Likewise, following analogous arguments, $\exists N_1$ such that, for any $n > N_1$ (not dependent on $u$):
\begin{align}
\mathbb{P}_{\rm MD}(u) = P\left(\frac{Z}{n} \leq \tau_n | \sigma_{\rm j}^2=u, H_1 \right) > 1 - \frac{\epsilon}{2}
\end{align}
for any $\tau_n > \sigma_{\rm w}^2 + u + \sigma_{\rm a}^2 + \delta$, as long as $u < c$.  Combining these results yields that for any $n > \mbox{max}(N_0,N_1)$:
\begin{align}
\mathbb{P}_{\rm FA}(u) + \mathbb{P}_{\rm MD}(u) \geq 1 - \frac{\epsilon}{2}
\end{align}
unless $\{u > c\}$ or $u \in \mathcal{A}=\{\sigma_{\rm w}^2 + u - \delta < \tau_n < \sigma_{\rm w}^2 + u + \sigma_{\rm a}^2 + \delta\}$.  Now,
\begin{align}
P(\mathcal{A}) & = P(\tau_n - \delta - \sigma_{\rm a}^2 - \sigma_{\rm w}^2 < U < \tau_n + \delta - \sigma_{\rm w}^2) \\
& \leq \frac{\sigma_{\rm a}^2 + 2 \delta}{\zeta}
\end{align}
where the last line follows by upper bounding the probability density function of $\sigma_{\rm j}^2$.  A choice of $\delta = \zeta \epsilon / 16$ and $\sigma_{\rm a}^2 = \zeta \epsilon / 8$ yields, via the Union Bound:
\begin{align}
P({\cal A}^c \cap \{\sigma_{\rm j}^2 \leq c\}) \geq 1 - \frac{\epsilon}{2}
\end{align}
and then the proof follows analogously to the end of that of Theorem 1.  This completes the proof for the case that the Alice-to-Willie channel is an AWGN channel.

Next, consider the case when the Alice-to-Willie channel is a $M=1$ block fading channel.  Let $\epsilon_2 > 0$ be the covertness constraint and set $\epsilon = \epsilon_2/2$.  Choose $\tilde{\sigma}_{\rm a}^2$ according to the AWGN case above such that Alice is covert if the average received power at Willie is $\tilde{\sigma}_{\rm a}^2$.  Finally, choose $P_{\rm f}$ such that: 

\begin{align}
P(\sigma_{\rm a}^2 < \tilde{\sigma}_{\rm a}^2) > 1 - \frac{\epsilon_2}{2}.
\end{align}  
Then, Alice can employ (constant) power $P_{\rm f}$ and satisfy the covertness constraint for any $\epsilon > 0$. \done



\subsection{\bf Proof of $\mathbf{o(n)}$ Covert Bits Transmitted for $\mathbf{M=1}$:}
\textit{
Consider the assumptions of the $M=1$ fading model and Alice's construction in Section \ref{sec:awgn} but with the jammer transmitting Gaussian noise drawn from a distribution with constant variance. If fading channels exist between all parties,  there exists a covert communication strategy s.t. Bob can reliably decode Alice's messages if she transmits $o(n)$ bits in $n$ channel uses.
}

\textbf{\textit{Proof:}}   By Theorem 2, Alice can transmit with $P_{\rm f} > 0$ not dependent on $n$ while remaining covert.  What remains is to demonstrate that Bob can decode the transmission with probability of error less than $\delta$ for any $\delta > 0$.  Conditioned on the fading variables $h^{\rm(a,b)}$, $h^{\rm(j,b)}$, the channel from Alice to Bob is an AWGN channel with signal-to-noise ratio:
\begin{equation}
\gamma = \frac{|h^{\rm (a,b)}|^2\frac{P_{\rm f}}{d^{\alpha}_{\rm a,b}}}{|h^{\rm (j,b)}|^2\frac{P_{\rm j}}{d^{\alpha}_{\rm j,b}}+\sigma_{\rm b}^2}.
\end{equation}
Hence, given the distributions of $h^{\rm(a,b)}$ and $h^{\rm(j,b)}$, there exists a constant rate $R$ such that the probability that $\gamma$ is large enough to support communication with reliability greater than $1 - \frac{\delta}{2}$ at rate $R$ is greater than $1 - \frac{\delta}{2}$ ($R$ is the $\frac{\delta}{2}$-outage capacity \cite{tse_vishwanath}, which is non-zero).  Since $o(n) < nR$ for all $n > N_0$ for some $N_0$, the result follows.
\done


\subsection{\bf Proof of Increasing $\mathbf{\Lambda(Z)}$ for the $M=1$ case for the Proof of Lemma 4:}
Let $\zeta = P_{\rm j} / d_{\rm j,w}^{\alpha}$.   Hence, in the fading model, the received jammer power $\sigma_{\mathrm{j}}^2$ is exponentially distributed with mean $\zeta$.  As in Section III, since the $t=0$ slot is the slot of interest, observations outside of $k=1,2,\ldots,n$ do not help Willie to detect a transmission by Alice in slot $t=0$.  Hence, it is sufficient to consider $\mathbf{Z}_0$ as the input to Willie's receiver.  As in Section III, we therefore suppress the slot index and denote Willie's observation by $\mathbf{Z} = [Z_1, Z_2, \ldots, Z_n]$.  It is then readily established that $Z=\sum_{i=1}^n|Z_i|^2$ is a sufficient statistic, with distribution under $H_0$ given by:
\begin{align}
f_{Z|H_0}(z | H_0) &= E_{\sigma_{\mathrm{j}}^2}\left[\left(\frac{1}{\pi(\sigma^2_{{\rm j}} + \sigma_{\mathrm{w}}^2)}\right)^{n}\exp\left(-\frac{z}{(\sigma^2_{{\rm j}} + \sigma^2_{\rm w})} \right)\right]\nonumber\\
&= \frac{1}{\pi^{n}}\int_0^{\infty} \left ( \frac{1}{u + \sigma_{\mathrm{w}}^2}\right)^{n} e^{-\frac{z}{(u+\sigma_{\mathrm{w}}^2)}}e^{-\frac{u}{\zeta}} du \nonumber \\
&=\frac{e^{\frac{\sigma_{\mathrm{w}}^2}{\zeta}}}{\pi^{n}}\int_{\sigma_{\mathrm{w}}^2}^{\infty} \left ( \frac{1}{v}\right)^{n} e^{-\frac{z}{v}} e^{-\frac{v}{\zeta}} dv. \label{eq:h0_fade_m1}
\end{align}
Via analogous arguments, the distribution when Alice transmits is:
\begin{align}
f_{Z|H_1}(z | H_1)&=\frac{e^{\frac{\sigma_{\mathrm{w}}^2+\sigma_{\mathrm{a}}^2}{\zeta}}}{\pi^{n}}\int_{\sigma_{\mathrm{w}}^2+\sigma_{\mathrm{a}}^2}^{\infty} \left ( \frac{1}{v}\right)^{n} e^{-\frac{z}{v}} e^{-\frac{v}{\zeta}} dv. \label{eq:h1_fade_m1}
\end{align}
Hence, in this case the optimal decision rule for Willie becomes:
\begin{align}
\Lambda(Z)&= \frac{e^{\frac{\sigma_{\mathrm{a}}^2}{\zeta}} \int_{\sigma_{\mathrm{w}}^2+\sigma_{\mathrm{a}}^2}^{\infty}  \left ( \frac{1}{v}\right)^{n} e^{-\frac{Z}{v}} e^{-v/\zeta} dv} {\int_{\sigma_{\mathrm{w}}^2}^{\infty} \left ( \frac{1}{v}\right)^{n} e^{-\frac{Z}{v}} e^{-\frac{v}{\zeta}} dv} \mathop{\gtrless}_{H_0}^{H_1} \gamma.  \label{test2}
\end{align}

\noindent Now, consider any observation $Z=z^{(0)}$ that falls on the boundary between the decision regions:
\begin{align}
\Lambda(z^{(0)})&=\frac{e^{\frac{\sigma_{\mathrm{a}}^2}{\zeta}} \int_{\sigma_{\mathrm{w}}^2+\sigma_{\mathrm{a}}^2}^{\infty}  \left( \frac{1}{v}\right)^{n} e^{-\frac{z^{(0)}}{v}} e^{-\frac{v}{\zeta}} dv}{\int_{\sigma_{\mathrm{w}}^2}^{\infty} \left( \frac{1}{v}\right)^{n} e^{-\frac{z^{(0)}}{v}} e^{-\frac{v}{\zeta}} dv}  = \gamma,\label{eq:lem4_1}
\end{align}

\noindent and consider the LRT when Willie observes $z^{(0)} + \Delta$:
\begin{align}
\Lambda(z^{(0)}+\Delta)&=\frac{e^{\frac{\sigma_{\mathrm{a}}^2}{\zeta}} \int_{\sigma_{\mathrm{w}}^2+\sigma_{\mathrm{a}}^2}^{\infty}  \left( \frac{1}{v}\right)^{n} e^{-\frac{(z^{(0)}+\Delta)}{v}} e^{-\frac{v}{\zeta}} dv}{\int_{\sigma_{\mathrm{w}}^2}^{\infty} \left( \frac{1}{v}\right)^{n} e^{-\frac{(z^{(0)}+\Delta)}{v}} e^{-\frac{v}{\zeta}} dv}.  \label{eq:lem4_2}
\end{align}
The common integration term in the numerator and denominator of (\ref{eq:lem4_2}) is extracted to yield:
\begin{align}
\Lambda(z^{(0)}+\Delta)&=\bigg[e^{\frac{\sigma_{\mathrm{a}}^2}{\zeta}} \int_{\sigma_{\mathrm{w}}^2+\sigma_{\mathrm{a}}^2}^{\infty}  \left( \frac{1}{v}\right)^{n} e^{-\frac{(z^{(0)}+\Delta)}{v}} e^{-\frac{v}{\zeta}} dv\bigg]\nonumber\\
&\cdot\bigg[\int_{\sigma_{\mathrm{w}}^2}^{\sigma_{\rm w}^2 + \sigma_{\rm a}^2} \left( \frac{1}{v}\right)^{n} e^{-\frac{(z^{(0)}+\Delta)}{v}} e^{-\frac{v}{\zeta}} dv \nonumber\\
&\qquad + \int_{\sigma_{\rm w}^2 + \sigma_{\rm a}^2}^{\infty} \left( \frac{1}{v}\right)^{n} e^{-\frac{(z^{(0)}+\Delta)}{v}} e^{-\frac{v}{\zeta}} dv\bigg]^{-1}.  \label{eq:lem4_3}
\end{align}


\noindent Next, (\ref{eq:lem4_3}) is normalized by the common integration range $\int_{\sigma_{\rm w}^2 + \sigma_{\rm a}^2}^{\infty} \left( \frac{1}{v}\right)^{n} e^{-\frac{(z^{(0)}+\Delta)}{v}} e^{-\frac{v}{\zeta}} dv$ to yield:
\begin{align}
\Lambda(z^{(0)}+\Delta)&=\frac{e^{\frac{\sigma_{\mathrm{a}}^2}{\zeta}} }{\frac{\int_{\sigma_{\mathrm{w}}^2}^{\sigma_{\rm w}^2 + \sigma_{\rm a}^2} \left( \frac{1}{v}\right)^{n} e^{-\frac{(z^{(0)}+\Delta)}{v}} e^{-\frac{v}{\zeta}} dv}{\int_{\sigma_{\rm w}^2 + \sigma_{\rm a}^2}^{\infty} \left( \frac{1}{v}\right)^{n} e^{-\frac{(z^{(0)}+\Delta)}{v}} e^{-\frac{v}{\zeta}} dv} + 1}.  \label{eq:lem4_4}
\end{align}

\noindent The Second Mean Value Theorem \cite[Chapter~4.7]{math_book_smvt} implies that $\exists c_1 \in (\sigma_{\mathrm{w}}^2, \sigma_{\mathrm{w}}^2+\sigma_{\mathrm{a}}^2)$ such that:
\begin{align}
e^{-\frac{\Delta}{c_1}}&\int_{\sigma_{\mathrm{w}}^2}^{\sigma_{\mathrm{w}}^2+\sigma_{\mathrm{a}}^2} \left(\frac{1}{v}\right)^{n}  e^{-\frac{z^{(0)}}{v}}e^{-\frac{v}{\zeta}}dv \nonumber\\
&\qquad\qquad=
\int_{\sigma_{\mathrm{w}}^2}^{\sigma_{\mathrm{w}}^2+\sigma_{\mathrm{a}}^2} \left(\frac{1}{v}\right)^{n} e^{-\frac{(z^{(0)}+\Delta)}{v}}e^{-\frac{v}{\zeta}}dv.\label{eq:lem4_5}
\end{align}

\noindent Similarly, because $e^{-\frac{\Delta}{\sigma_{\rm w}^2 + \sigma_{\rm a}^2}} \leq e^{-\frac{\Delta}{v}} \leq 1$ for $v \in [\sigma_{\rm w}^2 + \sigma_{\rm a}^2, \infty)$,
\begin{align}
&e^{-\frac{\Delta}{\sigma_{\rm w}^2 + \sigma_{\rm a}^2}}\int_{\sigma_{\mathrm{w}}^2+\sigma_{\mathrm{a}}^2}^{\infty} \left(\frac{1}{v}\right)^{n} e^{-\frac{z^{(0)}}{v}}e^{-\frac{v}{\zeta}}dv \nonumber\\
&\qquad\qquad\qquad\qquad  \leq
\int_{\sigma_{\mathrm{w}}^2+\sigma_{\mathrm{a}}^2}^{\infty} \left(\frac{1}{v}\right)^{n} e^{-\frac{(z^{(0)}+\Delta)}{v}}e^{-\frac{v}{\zeta}}dv \nonumber\\
&\qquad\qquad \qquad\qquad \leq \int_{\sigma_{\mathrm{w}}^2+\sigma_{\mathrm{a}}^2}^{\infty} \left(\frac{1}{v}\right)^{n} e^{-\frac{z^{(0)}}{v}}e^{-\frac{v}{\zeta}}dv \label{eq:lem4_6}
\end{align}
\noindent which implies:
\begin{align}
e^{-\frac{\Delta}{\sigma_{\rm w}^2 + \sigma_{\rm a}^2}} \leq \frac{\int_{\sigma_{\mathrm{w}}^2+\sigma_{\mathrm{a}}^2}^{\infty} \left(\frac{1}{v}\right)^{n} e^{-\frac{(z^{(0)}+\Delta)}{v}}e^{-\frac{v}{\zeta}}dv}{\int_{\sigma_{\mathrm{w}}^2+\sigma_{\mathrm{a}}^2}^{\infty} \left(\frac{1}{v}\right)^{n} e^{-\frac{z^{(0)}}{v}}e^{-\frac{v}{\zeta}}dv} \leq 1.  \label{eq:lem4_7}
\end{align}

\noindent Hence, the ratio of the integrals in (\ref{eq:lem4_7}) is either equal to one, or $\exists c_2 \in [\sigma_{\mathrm{w}}^2+\sigma_{\mathrm{a}}^2, \infty)$ such that:
\begin{align}
&e^{-\frac{\Delta}{c_2}}\int_{\sigma_{\mathrm{w}}^2  + \sigma_{\mathrm{a}}^2}^{\infty} \left(\frac{1}{v}\right)^{\frac{n}{M}} e^{-\frac{z^{(0)}}{v}}e^{-\frac{v}{\zeta}}dv \nonumber\\
&\qquad\qquad\quad =\int_{\sigma_{\mathrm{w}}^2+\sigma_{\mathrm{a}}^2}^{\infty} \left(\frac{1}{v}\right)^{\frac{n}{M}} e^{-\frac{(z^{(0)}+\Delta)}{v}}e^{-\frac{v}{\zeta}}dv.\label{eq:lem4_8}
\end{align}

\noindent If there exists such a $c_2 \in [\sigma_{\mathrm{w}}^2+\sigma_{\mathrm{a}}^2, \infty)$, then:
\begin{align}
\Lambda(z^{(0)}+\Delta)&=\frac{e^{\frac{\sigma_{\mathrm{a}}^2}{\zeta}} }{\frac{e^{-\frac{\Delta}{c_1}}\int_{\sigma_{\mathrm{w}}^2}^{\sigma_{\rm w}^2 + \sigma_{\rm a}^2} \left( \frac{1}{v}\right)^{n} e^{-\frac{z^{(0)}}{v}} e^{-\frac{v}{\zeta}} dv}{e^{-\frac{\Delta}{c_2}}\int_{\sigma_{\rm w}^2 + \sigma_{\rm a}^2}^{\infty} \left( \frac{1}{v}\right)^{n} e^{-\frac{z^{(0)}}{v}} e^{-\frac{v}{\zeta}} dv} + 1}  \label{eq:lem4_9}\\
&>\frac{e^{\frac{\sigma_{\mathrm{a}}^2}{\zeta}} }{\frac{\int_{\sigma_{\mathrm{w}}^2}^{\sigma_{\rm w}^2 + \sigma_{\rm a}^2} \left( \frac{1}{v}\right)^{n} e^{-\frac{z^{(0)}}{v}} e^{-\frac{v}{\zeta}} dv}{\int_{\sigma_{\rm w}^2 + \sigma_{\rm a}^2}^{\infty} \left( \frac{1}{v}\right)^{n} e^{-\frac{z^{(0)}}{v}} e^{-\frac{v}{\zeta}} dv} + 1}  \label{eq:lem4_10}
\end{align}
\noindent where (\ref{eq:lem4_10}) follows by noting that $e^{-\frac{\Delta}{x}}$ is monotonically increasing in $x$ and $c_2 > c_1$.  And (\ref{eq:lem4_10}) also holds if the ratio of the integrals in (\ref{eq:lem4_7}) is equal to one, in which case $e^{-\frac{\Delta}{c_2}}$ is replaced by 1 in (\ref{eq:lem4_9}).  Multiplying (\ref{eq:lem4_10}) through by the term $\int_{\sigma_{\rm w}^2 + \sigma_{\rm a}^2}^{\infty} \left( \frac{1}{v}\right)^{n} e^{-\frac{z^{(0)}}{v}} e^{-\frac{v}{\zeta}} dv$ yields:
\begin{align}
\Lambda(z^{(0)}+\Delta)&>\bigg[e^{\frac{\sigma_{\mathrm{a}}^2}{\zeta}} \int_{\sigma_{\rm w}^2 + \sigma_{\rm a}^2}^{\infty} \left( \frac{1}{v}\right)^{n} e^{-\frac{z^{(0)}}{v}} e^{-\frac{v}{\zeta}} dv\Bigg]\nonumber\\
&\quad\cdot\Bigg[\int_{\sigma_{\mathrm{w}}^2}^{\sigma_{\rm w}^2 + \sigma_{\rm a}^2} \left( \frac{1}{v}\right)^{n} e^{-\frac{z^{(0)}}{v}} e^{-\frac{v}{\zeta}} dv \nonumber\\
&\quad\, + \int_{\sigma_{\rm w}^2 + \sigma_{\rm a}^2}^{\infty} \left( \frac{1}{v}\right)^{n} e^{-\frac{z^{(0)}}{v}} e^{-\frac{v}{\zeta}} dv\Bigg]^{-1}  \label{eq:lem4_11}\\
&=\frac{e^{\frac{\sigma_{\mathrm{a}}^2}{\zeta}} \int_{\sigma_{\rm w}^2 + \sigma_{\rm a}^2}^{\infty} \left( \frac{1}{v}\right)^{n} e^{-\frac{z^{(0)}}{v}} e^{-\frac{v}{\zeta}} dv}{\int_{\sigma_{\rm w}^2}^{\infty} \left( \frac{1}{v}\right)^{n} e^{-\frac{z^{(0)}}{v}} e^{-\frac{v}{\zeta}} dv}  \label{eq:lem4_12}\\
&= \gamma \label{eq:lem4_13}
\end{align}
where (\ref{eq:lem4_13}) follows from the assumption in (\ref{eq:lem4_1}).  Hence, if an observation $z^{(0)}$ is such that $\Lambda(z^{(0)})=\gamma$, then an increase in the observed power $z$ results in $\Lambda(z)> \gamma$. $\done$

\subsection{\bf Proof of Lemma 5}

To bound the probability of $R^{\delta}_{\rm B}(n)$, we construct a set of $R^{\delta}_{{\rm B}_M}(n)$ that includes all points in $R^{\delta}_{\rm B}(n)$ and measure the probability of $R^{\delta}_{{\rm B}_M}(n)$.   
Define the $(M-1)$-dimensional vector $\mathbf{x}_{\sim m} = [x_1,x_2,\ldots,x_{m-1},x_{m+1},\ldots,x_M]$ as the vector $\bx$ with the $m^{\rm th}$ component removed. 
The set $R^{\delta}_{{\rm B}_M}(n)$ is then created iteratively as follows. 

For the initialization step, consider solving for the values (if there are any) of $x_1$, the first component of the vector $\bx = [x_1,x_2,\ldots,x_M]$, for which $\bx \in {\rm C}(n)$, with the other components fixed.  By Lemma \ref{lem:det_M>1}, we know that, for a given $[x_2, x_3, \ldots x_M]$, the set of $x_1$ such that $\mathbf{x} \in {\rm C}(n)$ consists of no points or a single point; thus, let:
\begin{align}
g({ \mathbf{x}_{\sim 1}}) = \left \{ \begin{array}{ll}
                                  \mbox{undefined},~ & \mbox{no}~x_1~\mbox{s.t.}~\mathbf{x} \in {\rm C}(n) \\
                                  x_1,~ & \mbox{a single}~x_1~\mbox{s.t.}~\mathbf{x} \in {\rm C}(n). 
            \end{array} \right .
\end{align}

\noindent Then, define:
\begin{align}
R^{\delta}_{{\rm B}_1}(n) = \{\bx: x_1 \in (g({ \mathbf{x}_{\sim 1}}) - \delta, g({ \mathbf{x}_{\sim 1}}) + \delta) \}.
\end{align} 
where it will be implicitly assumed that we only include $\bx$ for which $g({ \mathbf{x}_{\sim 1}})$ is defined.

We then start with $R^{\delta}_{{\rm B}_1}(n)$ and iterate in a similar fashion through the other dimensions to successively build $R^{\delta}_{{\rm B}_m}(n)$ from $R^{\delta}_{{\rm B}_{m-1}}(n), m=2,3,\ldots, M$, except now we are adding onto both sides of a region rather than a curve in each case.   As a consequence of Lemma 4, note that, as we fix all of the components except $x_m$ and then consider the $x_m$ s.th. $\bx$ falls in a given boundary region, we always get no solution, a single point, or an interval.
Hence, define
\begin{align}
l_m(\bx_{\sim m}) = \begin{cases}
\mbox{undefined},&\mbox{\parbox[t]{.15\textwidth}{no $x_m$ s.t. \\ $\bx\in R^{\delta}_{{\rm B}_{m-1}}(n)$}}\\
\inf \{x_m:\bx\in R^{\delta}_{{\rm B}_{m-1}}(n)\}, &\mbox{\parbox[t]{.15\textwidth}{$\exists~x_m$ s.t. \\$\;\bx\in R^{\delta}_{{\rm B}_{m-1}}(n)$}}
\end{cases}
\end{align}

and
\begin{align}
u_m(\bx_{\sim m}) = \begin{cases}
\mbox{undefined},&\mbox{\parbox[t]{.15\textwidth}{no $x_m$  s.t. \\ $\bx\in R^{\delta}_{{\rm B}_{m-1}}(n)$}}\\
\sup \{x_m:\bx\in R^{\delta}_{{\rm B}_{m-1}}(n)\}, &\mbox{\parbox[t]{.15\textwidth}{$\exists~x_m$ s.t. \\$\bx\in R^{\delta}_{{\rm B}_{m-1}}(n)$.}}
\end{cases}
\end{align}
We then construct $R^{\delta}_{{\rm B}_{m}}(n), m=2,3,\ldots,M$ from $R^{\delta}_{{\rm B}_{m-1}}(n)$  as follows:
\begin{align}
R^{\delta}_{{\rm B}_{m}}(n) &= R^{\delta}_{{\rm B}_{m-1}}(n)\nonumber\\
&\qquad~\bigcup~\bigg\{\bx: x_m\in \bigg(l_m(\bx_{\sim m})-\delta,~l_m(\bx_{\sim m})\bigg)\bigg\}\nonumber\\
&\qquad  \bigcup~\bigg\{\bx: x_m\in \bigg(u_m(\bx_{\sim m}),~u_m(\bx_{\sim m})+\delta\bigg)\bigg\}.
\label{union_three_things}
\end{align}
\noindent We are adding a layer of thickness $\delta$ to each side in dimension $m$ at the $m^{\th}$ stage.

Next we show that $R^{\delta}_{{\rm B}}(n)\subset R^{\delta}_{{\rm B}_{M}}(n)$.  By construction, $R^{\delta}_{{\rm B}_{1}}(n)$ contains all points $\bx$ such that $\exists~\by\in C(n)$ such that $|x_1-y_1|<\delta$ and $x_i=y_i, i=2,3,\ldots,M$.  Thus, by construction, $R^{\delta}_{{\rm B}_{2}}(n)$ contains all points $\bx$ such that $\exists\by\in C(n)$ such that $|x_1-y_1|<\delta$, $|x_2-y_2|<\delta$ and $x_i=y_i, i=3,4,\ldots,M$.  Continuing, $R^{\delta}_{{\rm B}_{M}}(n)$ contains all points $\bx$ such that $\exists\by\in C(n)$ such that $|x_1-y_1|<\delta$, $|x_2-y_2|<\delta,\ldots, |x_M-y_M|<\delta$.  Hence, $R^{\delta}_{{\rm B}}(n)\subset R^{\delta}_{{\rm B}_{M}}(n)$.

\begin{figure}[ht]
  \begin{center}
  \includegraphics[scale=.4]{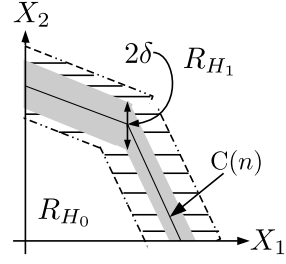}
  \end{center}
  \caption{An example showing the construction of the set $R^{\delta}_{{\rm B}_{M}}(n)$ that includes the boundary region for $M=2$ block fading conditions.  $X_1$ and $X_2$ are the normalized power measurements in the first and second block respectively.  The solid line ($-$) represents the boundary curve ${\rm C}(n)$.  The solid grey region represents the portion of the boundary that is defined by considering points that fall within $\delta$ of $C(n)$ in the first dimension.  The striped boundary region represents the portion of the boundary region constructed from iterating in the second dimension.}
  \label{fig:decision_regions_example}
\end{figure}

What remains is to measure the probability of $R^{\delta}_{{\rm B}_{M}}(n)$ by applying a union bound.  For $m=2,\ldots,M$, consider first the measure of the ``lower'' region added at the $m^{th}$ step.  Let
\begin{align}
L_m =  \bigg\{\bx: x_m\in \bigg(l_m(\bx_{\sim m})-\delta,~l_m(\bx_{\sim m})\bigg)\bigg\}
\end{align}
and note:
\begin{align}
&P( \boldsymbol{\sigma}^2_{\rm j} +\boldsymbol{\sigma}^2_{\rm w} \in L_m)\nonumber\\
&\qquad = \int_{L_m} \prod_{i=1}^M f_{\sigma^2_{{\rm j},i}+\sigma_{\rm w}^2}(x_i) dx_i \\
&\qquad \leq \int_{{\mathbf{x}_{\sim m}}} \int_{l({\mathbf{x}_{\sim m}}) -  \delta}^{l({\mathbf{x}_{\sim m}})}
 \prod_{i=1}^M f_{\sigma^2_{{\rm j},i}+\sigma_{\rm w}^2}(x_i) dx_i \\
&\qquad = \int_{{\mathbf{x}_{\sim m}}} \prod_{\stackrel{i=1}{i \neq m}}^M f_{\sigma_{{\rm j},i}^2 + \sigma_{\rm w}^2}(x_i) \nonumber\\ 
&\qquad\qquad \cdot\left [ \int_{l({\mathbf{x}_{\sim m}}) -  \delta}^{l({\mathbf{x}_{\sim m}})}
 f_{\sigma_{{\rm j},m}^2+\sigma_{\rm w}^2}(x_m) dx_m \right ] d({\mathbf{x}_{\sim m}}) \\
&\qquad \leq  \int_{{\mathbf{x}_{\sim m}}} \prod_{\stackrel{i=1}{i \neq m}}^M f_{\sigma_{{\rm j},i}^2 + \sigma_{\rm w}^2}(x_i) \nonumber\\
&\qquad\qquad\qquad\qquad\cdot~[\delta \sup_{x} f_{\sigma_{{\rm j},m}^2 + \sigma_{\rm w}^2}(x)]~ d({\mathbf{x}_{\sim m}}) \\
&\qquad = \delta \sup_{x} f_{\sigma_{{\rm j},1}^2 + \sigma_{\rm w}^2}(x). \label{last_step}
\end{align}
Likewise, defining
\begin{align}
U_m = \bigg\{\bx: x_m\in \bigg(u_m(\bx_{\sim m}),~u_m(\bx_{\sim m})+\delta\bigg)\bigg\},
\end{align}
it is shown by nearly identical steps that $P(U_m) \leq \delta \sup_{x} f_{\sigma_{{\rm j},1}^2 + \sigma_{\rm w}^2}(x)$ for $m=2,\ldots,M$.  Now, by the construction in (\ref{union_three_things}), a union bound implies that:
\begin{align}
P(R^{\delta}_{{\rm B}_{M}}(n)) & \leq P(R^{\delta}_{{\rm B}_{M-1}}(n)) + P(L_M) + P(U_M) \\
& \leq P(R^{\delta}_{{\rm B}_{M-1}}(n)) + 2 \delta \sup_{x} f_{\sigma_{{\rm j},1}^2 + \sigma_{\rm w}^2}(x).
\end{align}
 Repeating this argument for $M-1, M-2,\ldots,2$ and recognizing that $P(R^{\delta}_{{\rm B}_{1}}(n))$ can be bounded in a set of steps analogous to those leading up to (\ref{last_step}) yields: 
\begin{align}
P(R^{\delta}_{{\rm B}_{M}}(n)) \leq 2 M \delta \sup_{x} f_{\sigma_{{\rm j},1}^2 + \sigma_{\rm w}^2}(x).
\end{align}
Hence, a selection of $\delta = \epsilon / (2 M \sup_{x} f_{\sigma_{{\rm j},1}^2 + \sigma_{\rm w}^2}(x))$ yields $P(\mathbf{h}: \boldsymbol{\sigma}^2_{\rm j} +\boldsymbol{\sigma}^2_{\rm w} \in R^{\delta}_{\rm B_{M}}(n)) < \epsilon$ and thus $P(\mathbf{h}: \boldsymbol{\sigma}^2_{\rm j} +\boldsymbol{\sigma}^2_{\rm w} \in R^{\delta}_{\rm B}(n)) < \epsilon$. \done

\end{document}